\documentclass[aps,preprint,superscriptaddress,amssymb,citeautoscript]{revtex4}

\usepackage{graphicx}
\usepackage{dcolumn}
\usepackage{bm}
\usepackage{multirow}
\usepackage{amsmath}
\usepackage[colorlinks=true,linkcolor=blue,urlcolor=blue, citecolor=blue]{hyperref}
\usepackage{array}

\graphicspath{{./figures/}}

\begin{document}
\title{Light-tunable three-phase coexistence in mixed halide perovskites}
\author{Zehua Chen}
\affiliation{Materials Simulation and Modelling, Department of Applied Physics, Eindhoven University of Technology, 5600 MB Eindhoven, The Netherlands}
\affiliation{Center for Computational Energy Research, Department of Applied Physics, Eindhoven University of Technology, P.O. Box 513, 5600 MB Eindhoven, The Netherlands}

\author{Geert Brocks}
\affiliation{Materials Simulation and Modelling, Department of Applied Physics, Eindhoven University of Technology, 5600 MB Eindhoven, The Netherlands}
\affiliation{Center for Computational Energy Research, Department of Applied Physics, Eindhoven University of Technology, P.O. Box 513, 5600 MB Eindhoven, The Netherlands}
\affiliation{Computational Materials Science, Faculty of Science and Technology and MESA+ Institute for Nanotechnology, University of Twente, P.O. Box 217, 7500 AE Enschede, The Netherlands}

\author{Shuxia Tao}
\email{S.X.Tao@tue.nl}
\affiliation{Materials Simulation and Modelling, Department of Applied Physics, Eindhoven University of Technology, 5600 MB Eindhoven, The Netherlands}
\affiliation{Center for Computational Energy Research, Department of Applied Physics, Eindhoven University of Technology, P.O. Box 513, 5600 MB Eindhoven, The Netherlands}

\author{Peter A. Bobbert}
\email{P.A.Bobbert@tue.nl}
\affiliation{Center for Computational Energy Research, Department of Applied Physics, Eindhoven University of Technology, P.O. Box 513, 5600 MB Eindhoven, The Netherlands}
\affiliation{Molecular Materials and Nanosystems, Eindhoven University of Technology, P.O. Box 513, NL-5600 MB Eindhoven, The Netherlands}

\date{\today}

\begin{abstract}
\noindent
Mixed iodine-bromine perovskites used in solar cells undergo below a critical temperature an intrinsic demixing into phases with different iodine-bromine compositions. In addition, under illumination they show nucleation of an iodine-rich phase. We predict from thermodynamic considerations that in mixed iodine-bromine perovskites like MAPb(I$_{1-x}$Br$_x$)$_3$ the interplay of these effects can lead to coexistence of a bromine-rich, iodine-rich, and nearly iodine-pure nucleated phase. This three-phase coexistence occurs in a region in the composition-temperature phase diagram near the critical point for intrinsic demixing. We investigate the hysteresis in the evolution of this coexistence when temperature or illumination intensity are cycled. Depending on the particular way the coexistence is established, nearly iodine-pure nuclei should form either in the iodine-rich phase only or both in the bromine-rich and iodine-rich phases. Experimental verification of this fundamentally novel type of light-tunable three-phase coexistence should be possible by a combination of absorption and photoluminescence experiments.
\end{abstract}
\maketitle

\section{Introduction}

The interest in the fundamental physical processes of metal-halide perovskites is rapidly increasing because of their spectacular performance in photovoltaic and light emission applications \cite{Kojima2009,Green2014,Noel2014,Stranks2015, Jeon2015,McMeekin2016,Lin2018,Jung2019,AlAshouri2020}. One of these processes is halide segregation in mixed halide perovskites. Such perovskites have the unique feature that their band gap can be tuned by varying the halide concentration ratio. This tuneability can be exploited to optimize the optoelectronic performance of all-perovskite multi-junction solar cells \cite{Xiao2020,Wang2020} or tandem solar cells consisting of a silicon bottom cell and a perovskite top cell \cite{AlAshouri2020,Fu2022}. Halide segregation is a serious problem in such applications and fundamental understanding of this process is therefore of utmost importance.

Apart from its importance for developing efficient solar cells, halide segregation is a very interesting fundamental physical phenomenon in itself. The different halide ions can be viewed as particles of different substances that prefer to mix or demix, following fundamental thermodynamical laws. In particular, minimization of the Helmholtz free energy, consisting of an enthalpic and an entropic contribution, determines if a mixed or demixed state is preferred. The enthalpic contribution favours demixing while the entropic contribution favours mixing, and the competition between the two determines the preference for mixing or demixing. In the extensively studied mixed halide perovksite MAPb(I$_{1-x}$Br$_x$)$_3$ (MA stands for the organic cation methylammonium) this leads to a miscibility gap, i.e., a range of halide composition values $x_1<x<x_2$ for which there is no stable mixture \cite{Lehmann2019}. Such a miscibility gap had been predicted from a computational investigation within density functional theory combined
with a thermodynamic analysis \cite{Brivio2016}.
The resulting composition-temperature, $x$-$T$, phase diagram is similar to the well-known phase diagram of a binary mixture, where a binodal signals that the free energy of the demixed situation becomes lower than that of the mixed situation, and a spinodal signals the disappearance a free energy barrier between the mixed and demixed situation. The binodal and spinodal meet at a critical point ($x_{\rm c},T_{\rm c})$. Above the critical temperature $T_{\rm c}$ no demixing occurs because the entropic contribution to the free energy then always dominates.

What makes the halide segregation process especially interesting, and very different from demixing processes in ordinary binary mixtures, is the fact that it can be influenced by light. Even when the mixed state of a perovskite like MAPb(I$_{1-x}$Br$_x$)$_3$ is stable in the dark, low-band gap iodine-rich nuclei are formed under illumination \cite{Hoke2015,Beal2016,Draguta2017}. Since a return to the original mixed situation occurs when illumination is interrupted, the effect should have a thermodynamic origin. The effect was rationalized by funneling of photocarriers, generated by  illumination of the perovskite film, into these nuclei \cite{Draguta2017}. The nuclei originate from random composition fluctuations that tend to grow by inward diffusion of iodine ions and outward diffusion of bromine ions, a process that is driven by the energy gain of the photocarriers funneling into the low-band gap nuclei. When the illumination is interrupted, the driving force for the growth of the nuclei disappears and the fully mixed situation is gradually restored. A complete thermodynamic theory for the effect should not only consider the compositional free energy governing halide segregation in the dark, but also the free energy of the photocarriers redistributing by diffusion over the different phases. We recently made a first step in constructing such a theory, where the key element is the consideration of the {\em sum} of the compositional and photocarrier free energy, and minimization of this total free energy for a situation of two coexisting phases with different halide composition \cite{Chen2021}. When the total free energy of the two coexisting phases is lower than that of a single mixed phase, the demixed situation is preferred. With this analysis, we could explain various experimental observations on light-induced halide segregation \cite{Hoke2015,barker2017,Beal2016,McMeekin2016,Hutter2020,Braly2017,Rehman2017,Dang2019,bush2018}.

However, a major puzzle remained unsolved, namely the possible formation by halide segregation of {\em three} coexisting phases with different halide compositions. Such a situation could be compared to the well-known gas-liquid-solid coexistence occurring in many substances, like water, where the volume fraction of water molecules is different in the gas, liquid, and solid phase. Demonstration of such a three-phase coexistence would therefore not only shed further light on halide segregation in mixed halide perovskites, but also be of foundational interest.

An indication for such a three-phase coexistence is the finding in our previous analysis of a compositional and a light-induced binodal that meet at a `triple point' ($x_{\rm tr},T_{\rm tr}$), with $x_{\rm tr}$ and $T_{\rm tr}$ depending on the illumination intensity $I$ \cite{Chen2021}. The compositional binodal signals halide segregation into a bromine-rich and an iodine-rich phase, whereas the light-induced binodal signals the nucleation of a nearly iodine-pure phase from a mixed phase by the illumination. A three-phase coexistence of an iodine-rich, bromine-rich, and nearly iodine-pure phase at ($x_{\rm tr},T_{\rm tr}$) may thus be expected. However, if such three-phase coexistence would only occur at ($x_{\rm tr},T_{\rm tr}$), experimental observation of this phenomenon would be extremely difficult. It would require searching for a line in a three-dimensional ($x,T,I$) phase space, where changing $x$ would involve fabricating a range of perovskite samples with different iodine-bromine composition. On the other hand, if there would be a sufficiently large three-dimensional region in ($x,T,I$) phase space with three-phase coexistence, instead of only a line, experimental verification would be much easier. Because all present considerations of halide segregation are limited to two coexisting phases, this question could up to now not be addressed.

To address this question, we present in this work a thermodynamic theory for light-tunable halide segregation in mixed iodine-bromine perovskites that considers the possible coexistence of {\em three} phases. We apply our theory to MAPb(I$_{1-x}$Br$_x$)$_3$ and show that an intriguingly rich phase diagram is obtained, with two new types of binodal-spinodal combinations signalling different types of transitions from two-phase to three-phase coexistence. Most importantly, we show that an extended and experimentally accessible region of three-phase coexistence of an iodine-rich, bromine-rich, and nearly iodine-pure phase exists in ($x,T,I$) phase space near the critical point ($x_{\rm c},T_{\rm c}$) for intrinsic demixing in the dark.

\section{Methods}\label{section:methods}
The key element of the theory is the minimization of a free energy consisting of the sum over the different phases of a compositional and photocarrier free energy: 
\begin{equation}
\Delta F^\star\left(\{x\}, \{\phi\}, T\right) = \sum_{i}\phi_i\left[\Delta F(x_i, T) + n_i E_{\rm g}(x_i)\right].
\label{eq:free_three}
\end{equation}
Here, $\Delta F(x_i,T)$ is the compositional Helmholtz mixing free energy per formula unit (f.u.), $x_i$ the bromine concentration, and $\phi_i$ the volume fraction of phase $i=1,2,3$. Demanding local charge neutrality, the densities of photogenerated electrons and holes in each phase $i$ are both given by $n_i$, the photocarrier density per f.u. Since $n_i\ll 1$ we can use under equilibrium conditions Boltzmann statistics, so that we have
\begin{equation}
n_i/n_j=e^{-\left(E_{\rm g}(x_i)-E_{\rm g}(x_j)\right)/k_{\rm B}T},
\label{eq:n1_n2_n3}
\end{equation}
where $E_{\rm g}(x)$ is the band gap at Br concentration $x$ and $k_{\rm B}T$ the thermal energy. We assume that the diffusion length of the photocarriers is larger than the typical phase domain sizes, so that we can take the photocarrier density in each phase to be uniform. Neglecting the small differences in the volume $V$ per f.u.\ in the different phases, we have $\sum_{i} \phi_i  = 1$ and  $\sum_{i} \phi_ix_i = x$. In equilibrium, the total generation rate of photocarriers is equal to the total annihilation rate:
\begin{equation}
G=\sum_{i} \phi_i \left(n_i/\tau + k n_i^2/V\right),
\label{eq:steadystate}
\end{equation}
where we make the simplifying assumption that the generation rate per f.u.\ $G$ and the monomolecular and bimolecular recombination rates 1/$\tau$ and $k$ are the same in all phases. For the photocarrier generation rate in a thin perovskite film we take $G = I\alpha V/ h\nu$, where $I$ is the illumination intensity, $\alpha$ the absorption coefficient, and $h\nu$ the photon energy.

All results presented in this work are based on minimization of the total free energy $\Delta F^\star\left(\{x\}, \{\phi\}, T\right)$ in Eq.~(\ref{eq:free_three}) with respect to the $\phi_i$'s and $x_i$'s. The technical aspects of finding the free energy minima, establishing the character of these minima and and their appearance and disappearance, which determine the locations of the various binodals and spinodals in the $x$-$T$ phase diagram, are given in the Supporting Information. The related problems are fundamentally different from usual phase coexistence problems, where the free energy is the addition of the free energies of the different phases. This is not the case for the free energy Eq.~(\ref{eq:free_three}), because the photocarrier free energy $\phi_i n_i E_{\rm g}(x_i)$ of a certain phase $i$ depends on the volume fractions of the other phases via Eqs.~(\ref{eq:n1_n2_n3}) and (\ref{eq:steadystate}).

\section{Results}
\subsection{Thermodynamic theory for three-phase coexistence}

Our thermodynamic theory is based on a consideration of the minima of a total Helmholtz free energy that is the addition of a compositional and a photocarrier free energy; see  Section \ref{section:methods} (`Methods'). The compositional free energy is the sum of the compositional mixing free energies of the phases, labeled by $i$, with different bromine (Br) concentrations $x_i$, weighted with the volume fractions $\phi_i$ of each phase. The number of phases can be 1 (no demixing), 2 (demixing into two coexisting phases), or 3 (demixing into three coexisting phases). Regarding the calculation of the compositional free energy, we follow the procedure outlined in Ref.~\cite{Brivio2016}, as we also did in our previous analysis \cite{Chen2021}. Considering exclusively the compositional free energy, only two-phase coexistence can occur. A phase diagram similar to that of an ordinary binary mixture is then obtained, with a binodal and a spinodal \cite{Brivio2016}. Figure~\ref{fig:fig4} in the Appendix \ref{section:compositional_free_energy} reproduces the compositional mixing free energy $\Delta F(x,T)$ of MAPb(I$_{1-x}$Br$_x$)$_3$ calculated in Ref.~\cite{Chen2021} for different temperatures $T$. We will use this free energy also in the present work.

Under illumination, photocarriers will be generated, leading to a photocarrier free energy in each phase that should be added to the compositional free energy. Under charge-neutrality conditions, the densities of photogenerated electrons and holes are equal and given by a photocarrier density $n_i$ per perovskite formula unit (f.u.). The photocarrier free energy of each phase is then given by $\phi_i n_i E_{\rm g}(x_i)$, where $E_{\rm g}(x_i)$ is the band gap of phase $i$, which depends on the Br concentration $x_i$ of the phase. The dependence of the band gap of MAPb(I$_{1-x}$Br$_x$)$_3$ on $x$ has been experimentally determined and can be well described by the function $E_{\rm g}(x)=1.57 + 0.39 x + 0.33x^2$ eV \cite{Noh2013}, interpolating between the band gap of 1.57 eV for MAPbI$_3$ and 2.29 eV for MAPbBr$_3$. We assume a phase-independent photocarrier generation rate $G$ per f.u.\ that is for a thin perovskite film given by $G = I\alpha V/ h\nu$, where $I$ is the illumination intensity, $\alpha$ the absorption coefficient, $V$ the volume per f.u., and $h\nu$ the photon energy. Photocarriers recombine by monomolecular and bimolecular recombination, given by rates $1/\tau$ and $k$, respectively, which are assumed to be independent of the phase. The diffusion lengths of photocarriers are very long in perovskites and expected to be larger than the feature sizes of the different phase domains. An equilibrium photocarrier distribution over the different phases is therefore assumed obeying Boltzmann statistics. Since in practice $n_i\ll 1$, we can neglect state filling effects and use the Boltzmann factor $\exp(-E_{\rm g}/k_{\rm B}T)$ in the statistical weighing, where $k_{\rm B}$ is Boltzmann's constant. It is the photocarrier contribution to the free energy that leads to the light-induced nucleation of a nearly iodine-pure phase from a mixed phase, allowing a free energy lowering by funneling of photocarriers from the mixed phase to the low-band gap nuclei \cite{Draguta2017,Chen2021}.

Figure \ref{fig:fig1} illustrates the main findings of the present work when considering minimization of the sum of the compositional and photocarrier free energy in the presence of illumination when allowing for the coexistence of three phases. As in the dark, a decomposition into a Br-rich and an I-rich phase can take place, governed by the compositional free energy. At the same time, a lowering of the photocarrier free energy drives the light-induced formation of nearly I-pure nuclei. We will show that the interplay of these processes gives rise to two types of three-phase coexistence, with the nearly I-pure nuclei being present in (I) both the I-rich and B-rich phase, or (II) only the I-rich phase.

\subsection{Phase diagrams of MAPb(I$_{1-x}$Br$_x$)$_3$}

We consider the phase diagrams of MAPb(I$_{1-x}$Br$_x$)$_3$ for different illumination intensities $I$. Binodals and spinodals are found by determining the minima of the total free energy. As usual, a binodal signals the crossing of the free energies of two local minima with the lowest free energy and therefore a change of the character of the global free energy minimum. A spinodal signals the appearance or disappearance of a local minimum and hence the vanishing of a free energy barrier between two local minima. Determining free energy minima for the case of three-phase coexistence is much more complicated than for the case of two-phase coexistence and required the development of sophisticated search procedures; see the Appendices \ref{seciton:bin_spin_1_to_2_coexistence}-\ref{seciton:bin_spin_2_to_3_coexistence}. We take the values $\tau = 100$ ns and $k=10^{-10}$ cm$^3$ s$^{-1}$, applicable for a standard MAPbI$_3$ film \cite{Johnston2016}, $V=2.5\times10^{-22}$ cm$^3$/f.u.\ \cite{Chen2021}, $\alpha = 10^5$ cm$^{-1}$, and $h\nu = 3$ eV \cite{Draguta2017}.

Figures~\ref{fig:fig2}(a)-(c) show the phase diagrams of MAPb(I$_{1-x}$Br$_x$)$_3$ for illumination intensities $I$ corresponding to 2, 20, and 200 Sun ($I=100$ mW cm$^{-2}=1$ Sun), obtained from our previous analysis, which assumed possible coexistence of not more than two phases \cite{Chen2021}.
The blue and green curves indicate the compositional and light-induced binodals, respectively, where the mixed situation (white) becomes metastable for a transition to two-phase coexistence (grey). The red curve indicates the spinodal, where the free energy barrier between the mixed and two-phase coexistence situations disappears. In our previous analysis we suggested that at the junction between the compositional and light-induced binodal (arrow in Fig.~\ref{fig:fig2}(a)) three-phase coexistence might occur \cite{Chen2021}.

Figures~\ref{fig:fig2}(d)-(f) show the phase diagrams when possible coexistence of three phases is taken into account. Next to the binodals in Figs.~\ref{fig:fig2}(a)-(c), two new binodals (orange curves) appear, signalling that the free energy minimum of a three-phase coexistence as illustrated in Fig.~\ref{fig:fig1} becomes lower than that of two-phase coexistence. Because there are two types of two-phase coexistence, there are also two types of binodals, signalling a transition from (1) two-phase coexistence of nuclei of a nearly I-pure phase and an I-Br mixed phase to three-phase coexistence, and (2) two-phase coexistence of an I-rich and Br-rich phase to three-phase coexistence. Each of the two binodals is accompanied by a spinodal that signals the disappearance of the free energy barrier between the corresponding two-phase and three-phase coexistence minima. Figures~\ref{fig:fig2}(g)-(i) zoom in to the three-phase coexistence regions, where the pairs of binodals and spinodals are labeled accordingly by `1' and `2'. The spinodals continue in the two-phase coexistence regions, where they signal the disappearance of free energy barriers between the two types of two-phase coexistence (not shown). The main conclusion from the phase diagrams is that a sizeable three-phase coexistence region exists in ($x,T,I$) phase space near the critical point ($x_c,T_c$) for intrinsic halide demixing in the dark.

\subsection{Analysis of the three-phase coexistence}

We observe in Figs.~\ref{fig:fig2}(e), (f), (h), and (i) that the two types of spinodals may cross. In an experiment investigating the three-phase coexistence, two situations may therefore occur when decreasing or increasing $T$, indicated by the two double-headed arrows at $x=0.395$ and $0.435$ in Fig.~\ref{fig:fig2}(h). Figures \ref{fig:fig3}(a) and (b) show, for these two compositions and an illumination intensity $I=20$ Sun, results for the volume fractions $\phi_i$ and Br concentrations $x_i$ of the three phases, and the photocarrier fractions $f_i=\phi_i n_i/\sum_i \phi_i n_i$ in these phases. In plotting these results it is assumed that, after crossing a binodal when decreasing or increasing $T$, the system makes a transition to three-phase coexistence only after crossing the corresponding spinodal, because only then the free energy barrier for this transition has disappeared. The resulting supercooling or superheating leads to the hysteresis in the plots. We observe that volume fractions of about 0.03\% of the nearly I-pure phase can be reached, with a maximum Br concentration of about 0.45\% and a photocarrier fraction of $20$-$30$\%. The fact that the very small volume fraction of the nearly I-pure phase contains a relatively large fraction of the photocarriers is a consequence of the very strong effect of funneling of photocarriers into this phase. 
The fundamental difference between the two situations depicted in Figs.~\ref{fig:fig3}(a) and (b) is that in Fig.~\ref{fig:fig3}(a) a temperature range (pink) exists where the transition to the three-phase coexistence always occurs, while in Fig.~\ref{fig:fig3}(b) the transition requires overcoming a free energy barrier. The three-phase coexistence is then both for decreasing and increasing $T$ `hidden' behind this barrier, as indicated by the dotted lines in Fig.~\ref{fig:fig3}(b).

\begin{figure}[h]
\centering
\includegraphics[width=0.8\columnwidth]{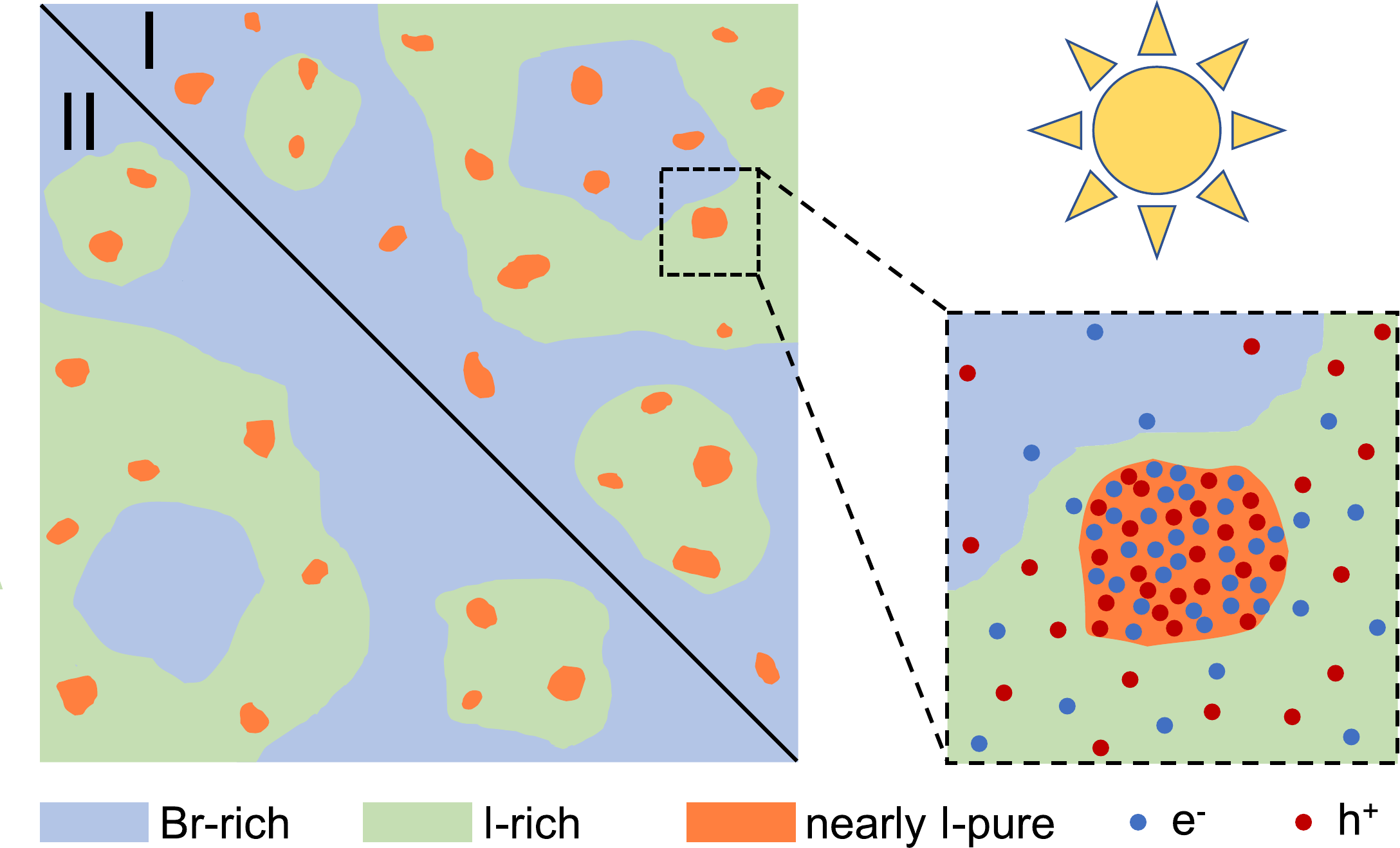}
\caption{\label{fig:fig1} Light-tunable three-phase coexistence in mixed I-Br perovskites. Blue: Br-rich phase. Green: I-rich phase. Orange: nearly I-pure nuclei. The nearly I-pure nuclei can be present in (I) both the I-rich and Br-rich phase or (II) only in the I-rich phase. Magnification: funneling of photogenerated electrons and holes into the low-band gap nearly I-pure nuclei, reducing their free energy.}
\end{figure}

The situation depicted as `I' in Fig.~\ref{fig:fig1}, with nearly I-pure nuclei in both the I-rich and Br-rich phase, can only arise when entering the three-phase coexistence region by decreasing $T$. The mixed I-Br phase in which the nearly I-pure nuclei are embedded then decomposes into an I-rich and a Br-rich phase; see the Appendix \ref{seciton:bin_spin_2_to_3_coexistence}. By contrast, when entering the three-phase coexistence region by increasing $T$, the nearly I-pure nuclei are exclusively generated in the I-rich phase (situation `II' in Fig.~\ref{fig:fig1}).

In between each binodal and its corresponding spinodal, the barrier for the formation of a third phase is determined by the free energy required to generate a critical nucleus of the third phase in either of the two existing phases. This barrier is, in the absence of defects and impurities, governed by an interface free energy. Determination of this interface free energy is beyond the scope of the present work. In practice defects and impurities might dominate the nucleation process, preventing supercooling and superheating. In that case the binodals and not the spinodals delineate the region where three-phase coexistence will be found experimentally.

We conclude by showing in Fig.~\ref{fig:fig3}(c) similar results as in Figs.~\ref{fig:fig3}(a) and (b), but now for varying illumination intensity $I$ for $T=252$ K and $x=0.395$ (black squares in Figs.~\ref{fig:fig2}(g)-(i)). A large hysteresis loop is observed that starts at 8 Sun, featuring `superillumination' until 133 Sun, where the nearly I-pure phase spontaneously appears with a volume fraction of about 0.08\% and Br concentration of about 0.2\%, containing about 13\% of the photocarriers. The same remarks as above about a free energy barrier to form a critical nucleus for the nearly I-pure phase apply to this situation. As is clear from Figs.~\ref{fig:fig2}(g)-(i), the way in which the three-phase coexistence region is entered for increasing $I$ is the same as for increasing $T$. This means that the nearly I-pure nuclei are in this case exclusively generated in the I-rich phase (situation `II' in Fig.~\ref{fig:fig1}).

\section{Summary, Conclusions, and Outlook}

Although we focused on a particular mixed halide perovskite,  MAPb(I$_{1-x}$Br$_x$)$_3$, and used particular parameter values, we expect that the concepts developed in this work will be applicable to mixed halide perovskites in general. We stress that substantial uncertainties in our quantitative predictions are expected, because we had to make several simplifying assumptions. Quantitative improvements of our work in several directions are possible. Accounting for different monomolecular and bimolecular recombination rates in the different phases will be relatively easy, but the information required for this is presently not available to us. Accounting for non-uniformity of the photocarrier distribution inside the phase domains will be more complicated, requiring information about the diffusion lengths of the photocarriers in comparison to the sizes and shapes of the domains. We further remark that we have applied thermodynamic equilibrium arguments, which cannot account for kinetic effects. One such effect is the kinetic trapping of Br ions inside the nearly I-pure nuclei \cite{Ruth2018}, possibly explaining why the observed Br concentration of $x\approx 0.2$ in these nuclei \cite{Hoke2015} is not as small as expected from equilibrium arguments. Finally, when changing temperature, perovskites may undergo structural phase transitions and volume changes that will interfere with halide segregation. Nevertheless, we expect the three-phase coexistence to be a robust phenomenon, also present in quantitatively improved versions of our analysis. In particular, we expect that our conclusion that the light-tunable three-phase coexistence should occur close to the critical point for intrinsic halide segregation in the dark will be unaffected by quantitative improvements.

We propose experimental verification of the predicted three-phase coexistence by a combination of absorption and photoluminescence experiments. Presence of both the I-rich and Br-rich phase can be detected by the occurrence of two step-like increases at the respective band gaps in the absorption spectrum. The additional nearly I-pure phase can be detected by a peak in the photoluminescence spectrum close to the band gap of MAPbI$_3$. Because of the funneling effect, a sizable peak should be visible, despite the very small volume fraction of this phase. We remark that extreme illumination intensities are not needed. We studied three-phase coexistence in Figs.~\ref{fig:fig3}(a) and (b) for $I=20$ Sun, but it is also present at $I=2$ Sun (see Figs.~\ref{fig:fig2}(d) and (g)) and even lower illumination intensities.

A worry could be that experimental verification requires cooling down to temperatures around $T_{\rm c}$, where slowing down of halide diffusion might hamper observation. Estimates of the diffusion barriers for vacancy-mediated diffusion of I and Br in MAPb(I$_{1-x}$Br$_x$)$_3$ are 0.17-0.25 and 0.23-0.43 eV, respectively \cite{brennan2018}. Taking 0.25 eV as a typical barrier and an Arrhenius $T$ dependence, we find that the diffusion speed is slowed down by a factor of about 3 at the calculated $T_{\rm c}=266$ K \cite{Chen2021} as compared to room temperature. Since light-induced halide demixing at room temperature occurs on time scales of hours \cite{Elmelund2020}, establishing equilibrium conditions for three-phase coexistence should still occur on experimentally feasible time scales. MAPb(I$_{1-x}$Br$_x$)$_3$ seems a good candidate for experimental verification, since the calculated $T_{\rm c}$'s of other AB(I$_{1-x}$Br$_x$)$_3$ perovskites are lower \cite{Chen2021}. We mention in this context that the precise value of the calculated $T_{\rm c}$ depends rather sensitively on technical details of the calculations. For example, in Ref.~\cite{Brivio2016} a value $T_{\rm c}=343$ K was reported for MAPb(I$_{1-x}$Br$_x$)$_3$, which is even above room temperature. Considerable uncertainties in calculated values of $T_{\rm c}$ should therefore be expected.

Experimental verification of the three phase coexistence would establish a fundamentally novel type of three-phase coexistence and provide a new viewing angle on demixing processes in mixed halide perovskites. The tunability by light makes the present three-phase coexistence fundamentally different from gas-liquid-solid coexistence as occurring, e.g., at the triple point of water. The insight that light can be used as a parameter to influence segregation processes can possibly also be applicable to other compound semiconductors where the band gap changes with composition.

\section{acknowledgements}
Z.C. acknowledges funding from Eindhoven University of Technology. S.T. acknowledges funding by the Computational Sciences for Energy Research (CSER) tenure track program of Shell and NWO (Project No. 15CST04-2) as well as NWO START-UP from the Netherlands.

\appendix
\section{Compositional mixing free energy}\label{section:compositional_free_energy}
In Fig.~\ref{fig:fig4} we reproduce the compositional mixing free energy $\Delta F(x,T)$ calculated in Ref.~\cite{Chen2021} that enters Eq.~(\ref{eq:free_three}) in `Methods' of the main text. This mixing free energy was calculated following the procedure outlined in Ref.~\cite{Brivio2016}, using density functional theory
with a thermodynamic analysis performed within the generalized quasi-chemical
approximation \cite{Sher1987}.

\section{Binodals and spinodals for mixed phase to two-phase coexistence}\label{seciton:bin_spin_1_to_2_coexistence}
We apply Eqs.~(\ref{eq:free_three})-(\ref{eq:steadystate}) in `Methods' of the main text to the case of halide segregation into two phases. Substituting $n_1$ and $n_2$, which can be obtained by solving Eqs.~(\ref{eq:n1_n2_n3}) and (\ref{eq:steadystate}) in the main text, Eq.~(\ref{eq:free_three}) in the main text becomes
\begin{equation}
\begin{aligned}
\Delta & F^\star (x_1, x_2, \phi_1, \phi_2, T) =  \phi_1 \Delta F(x_1) + \phi_2 \Delta F(x_2)\\
 &+ 2G\tau \frac{\phi_1 f(x_1) + \phi_2 f(x_2)}{\phi_1g(x_1) + \phi_2g(x_2) + \sqrt{(\phi_1g(x_1) + \phi_2g(x_2))^2 + D(\phi_1g^2(x_1) + \phi_2g^2(x_2))}},
\end{aligned}
\label{eq:totfree_two} 
\end{equation}
with 
\begin{equation}
D = 4Gk\tau^2/V,
\end{equation}
and 
\begin{equation}
\begin{aligned}
f(x) =& E_{\mathrm g}(x)e^{-E_{\mathrm g}(x)/k_{\mathrm B}T},\\
g(x) =& e^{-E_{\mathrm g}(x)/k_{\mathrm B}T}.
\end{aligned}
\label{eq:fx_gx}
\end{equation}

The $T$-dependence in Eq.~(\ref{eq:totfree_two}) has been suppressed. We should minimize Eq.~(\ref{eq:totfree_two}) under the conditions $\phi_1+\phi_2=1$ and $\phi_1x_1+\phi_2x_2=x$.

To find the binodals in the $x$-$T$ phase diagram, we investigate the possibility, starting from the mixed situation with Br concentration $x$, to decrease the free energy by demixing through nucleation of a phase with a Br concentration $x_2\neq x$ with a small volume fraction $\delta\phi\equiv \phi_2$. The free energy, Eq.~(\ref{eq:totfree_two}), in the mixed situation is $\Delta F^\star(x, x, 1, 0, T)$, while the free energy in the demixed situation is $\Delta F^\star(x_1, x_2, 1-\delta\phi, \delta\phi, T)$, with, to linear order in $\delta\phi$, $x_1 = x-(x_2-x)\delta\phi$. The difference in free energy between the demixed and mixed situations is to linear order in $\delta\phi$:
\begin{equation}
\begin{aligned}
\delta & \Delta F^\star =  \delta \phi \bigg\{\Delta F(x_2) - \Delta F(x) + (x - x_2)\Delta F^\prime(x) \\
&+ 2G\tau \bigg[ \frac{f(x_2) - f(x) + (x - x_2)f^\prime(x)}{(1 + \sqrt{1 + D}) g(x)} \\
&- \frac{f(x)}{((1 + \sqrt{1 + D}) g(x))^2} \bigg(g(x_2) - g(x) + (x - x_2)g^\prime(x)\\
& + \frac{2g(x)(g(x_2) - g(x) + (x-x_2)g^\prime(x)) + D(g^2(x_2) -g^2(x) + 2(x-x_2)g(x)g^\prime(x))}{2\sqrt{1+D}g(x)}\bigg)\bigg]\bigg\}.
\end{aligned}
\label{eq:2phasebinodal}
\end{equation}
When $\delta \Delta F^\star < 0$, the demixed situation has a lower free energy than the mixed situation. We therefore find the binodals in Figs.~\ref{fig:fig2}(a)-(c) of the main text for a certain illumination intensity $I$ by looking if a value $x_2$ of a nucleated phase exists for which $\delta \Delta F^\star =  0$.

To find the spinodals we consider the possibility to decrease the free energy by generating a volume fraction $\phi$ of a phase with a slightly different concentration $x_2=x+\delta x$. The free energy in the demixed situation can be written as $\Delta F^\star(x-\phi\delta x/(1-\phi), x + \delta x, 1-\phi, \phi,T)$. To second order in $\delta x$ the difference in free energy becomes
\begin{equation}
\begin{aligned}
\delta \Delta F^\star = &  \frac{\phi(\delta x)^2}{2(1-\phi)} \bigg\{\Delta F^{\prime\prime}(x) + 2G\tau\bigg[ \frac{f^{\prime\prime}(x)}{(1 + \sqrt{1 + D}) g(x)} - \frac{f(x)}{((1 + \sqrt{1 + D}) g(x))^2} \bigg(g^{\prime\prime}(x)\\
& + \frac{g(x)g^{\prime\prime}(x) + D([g^\prime(x)]^2 + g(x)g^{\prime\prime}(x))}{\sqrt{1+D}g(x)}\bigg)\bigg]\bigg\}.
\end{aligned}
\label{eq:2phasespinodal}
\end{equation}
Putting $\delta \Delta F^\star =  0$ yields the spinodal separating the unstable from the metastable region in the $x$-$T$ phase diagrams shown in Figs.~\ref{fig:fig2}(a)-(c) of the main text. 

We remark that Eqs.~(\ref{eq:2phasebinodal}) and (\ref{eq:2phasespinodal}) are more accurate than Eqs.~(11) and (13) in Ref.~\cite{Chen2021}. Equations~(\ref{eq:2phasebinodal}) and (\ref{eq:2phasespinodal}) consider  bimolecular recombination in both the parent and nucleated phases while Eqs.~(11) and (13) in Ref.~\cite{Chen2021} neglect bimolecular recombination in the parent phase. For the illumination intensities studied in Ref.~\cite{Chen2021} this is a good approximation, but in the present work this not a good approximation anymore for an illumination intensity of 200 Sun.

\section{Extension of common tangent method}\label{seciton:extension_common_tangent}
The binodals also follow from demanding that $\partial \Delta F^\star/\partial x_2 = 0$ and $\partial \Delta F^\star/\partial \phi_2 = 0$ under the conditions $\phi_1 + \phi_2 =1$ and $\phi_1x_1 + \phi_2x_2 = x$, with  $\Delta F^\star$ given by Eq.~(\ref{eq:totfree_two}). When only considering the compositional free energy this leads to the common tangent equations 
$\Delta F(x_2) =  \Delta F(x_1) - (x_1 - x_2)\Delta F^\prime(x_1)$ and $\Delta F^\prime(x_2)  = \Delta F^\prime(x_1)$. With the inclusion of the photocarrier free energy we obtain instead
\begin{equation}
\begin{aligned}
\Delta F(x_2) = & \Delta F(x_1) - (x_1 - x_2)\Delta F^\prime(x_1) - 2G\tau \bigg[ \frac{f(x_2) - f(x_1) + (x_1 - x_2)f^\prime(x_1)}{\phi_1 g(x_1) + \phi_2 g(x_2) + H_0}\\
&- \frac{\phi_1 f(x_1) + \phi_2 f(x_2)}{(\phi_1g(x_1) + \phi_2 g(x_2) + H_0)^2}\bigg(g(x_2) - g(x_1) + (x_1 - x_2)g^\prime(x_1) + \frac{H_1}{2H_0}\bigg)\bigg],
\end{aligned}
\label{eq:dF_dphi2}
\end{equation}
and 
\begin{equation}
\begin{aligned}
\Delta F^\prime(x_2)  = &\Delta F^\prime(x_1) - 2G\tau \bigg [ \frac{f^\prime(x_2)-f^\prime(x_1)}{\phi_1 g(x_1) + \phi_2 g(x_2) + H_0}\\
&-\frac{\phi_1 f(x_1) + \phi_2 f(x_2)}{(\phi_1g(x_1) + \phi_2 g(x_2) + H_0)^2}\bigg(g^\prime(x_2) - g^\prime(x_1) + \frac{H_2}{2H_0}\bigg) \bigg],
\end{aligned}
\label{eq:dF_dx2}
\end{equation}
where $H_0$, $H_1$ and $H_2$ are
\begin{equation}
\begin{aligned}
H_0 =& \sqrt{(\phi_1g(x_1) + \phi_2g(x_2))^2 + D(\phi_1g^2(x_1) + \phi_2g^2(x_2))},\\
H_1 =& 2(\phi_1g(x_1) + \phi_2 g(x_2))(g(x_2) - g(x_1) + (x_1 - x_2)g^\prime(x_1)) \\
&+ D(g^2(x_2) - g^2(x_1) + 2g(x_1)g^\prime(x_1)(x_1 - x_2)),\\
H_2 =& 2(\phi_1g(x_1) + \phi_2g(x_2))(g^\prime(x_2) -g^\prime(x_1)) + 2D(g(x_2)g^\prime(x_2) -g(x_1)g^\prime(x_1)).
\end{aligned}
\end{equation} 
Putting $\phi_1 = 1 - \delta\phi$, $\phi_2 = \delta\phi$ and $x_1 = x-(x_2-x)\delta\phi$ in Eq.~(\ref{eq:dF_dphi2}) and linearizing in $\delta \phi$ yields Eq.~(\ref{eq:2phasebinodal}).

\section{Binodals and spinodals for two-phase to three-phase coexistence}\label{seciton:bin_spin_2_to_3_coexistence}
\begin{figure}[h]
\centering
\includegraphics[width=0.8\columnwidth]{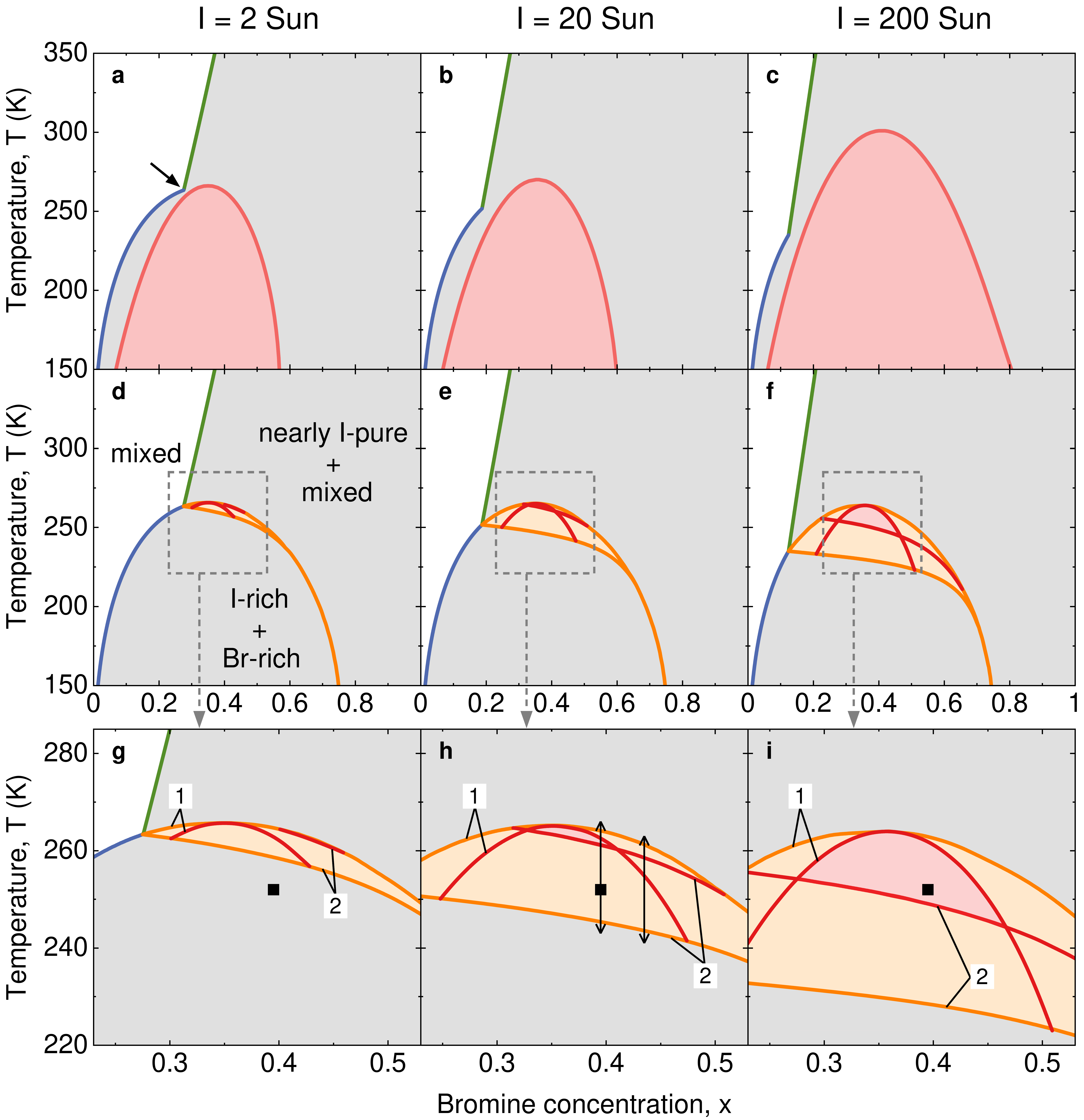}
\caption{\label{fig:fig2} Phase diagrams of MAPb(I$_{1-x}$Br$_x$)$_3$. (a)-(c) Phase diagrams of MAPb(I$_{1-x}$Br$_x$)$_3$ for different illumination intensities when accounting for coexistence of not more than two phases. Blue curve: compositional binodal. Green curve: light-induced binodal. Red curve: spinodal. Arrow in a: suggested `triple point' for three-phase coexistence. (d)-(f) Phase diagrams when accounting for coexistence of three phases. Orange curves: binodals for transitions from the two types of two-phase coexistence to three-phase coexistence. Red curves: corresponding spinodals. Yellow (pink) region: two-phase coexistence partly (fully) unstable for formation of third phase. (g)-(i) Magnifications of three-phase coexistence regions with labeling of binodal-spinodal pairs for the two types of transitions. Double-headed arrows in (h): temperature ranges and Br concentrations $x$ in Figs.~\ref{fig:fig3}(a) (left arrow) and (b) (right arrow). Black squares in (g)-(i): temperature $T$ and Br concentration $x$ in Fig.~\ref{fig:fig3}(c).}
\end{figure}

To determine the binodals for two-phase to three-phase coexistence, we start with finding the minimum of the two-phase free energy in the metastable (grey) and unstable (pink) regions of the phase diagrams Figs.~\ref{fig:fig2}(a)-(c) of the main text using a standard minimization algorithm. From this, the free energies $\Delta F^{\star}$, volume fractions $\phi_1$ and $\phi_2$, and Br concentrations $x_1$ and $x_2$ of the local and global minima are determined. The comparison of the free energies of the global minima leads to two distinct two-phase coexistence regions, separated by purple lines, as shown in Fig.~\ref{fig:fig5} for MAPb(I$_{1-x}$Br$_{x}$)$_3$ at different illumination intensities. In the upper right region we have `nearly I-pure + mixed' coexistence and in the lower left region `I-rich + Br-rich' coexistence.

To determine the binodals we consider the possibility, starting from the free energy of the global two-phase coexistence minimum, to further decrease the free energy by generating a third phase with Br concentration $x_3\neq x_1, x_2$ with a small volume fraction $\delta\phi\equiv \phi_3$. The two-phase free energy is obtained from Eq.~(\ref{eq:totfree_two}) and the three-phase free energy is $\Delta F^\star(x_1 + c_1\delta \phi, x_2 + c_2\delta \phi, x_3, \phi_1 - c\delta\phi, \phi_2 - (1-c)\delta \phi, \delta\phi, T)$, with, to linear order in $\delta\phi$, 
\begin{equation}
c_1\phi_1 + c_2\phi_2 - cx_1 - (1-c)x_2 + x_3 = 0,
\label{eq:x3}
\end{equation}
where $c_1$, $c_2$ and $c$ are constants. 
The free energy difference between the three-phase and two-phase coexistence is to linear order in $\delta\phi$:
\begin{equation}
\begin{aligned}
\delta \Delta F^ \star =&\delta \phi \bigg\{ \left(c_1\phi_1\Delta F^\prime(x_1) + c_2\phi_2\Delta F^\prime(x_2) - c\Delta F(x_1) - (1-c)\Delta F(x_2) + \Delta F(x_3) \right)\\
&+ 2G\tau \bigg[ \frac{c_1\phi_1 f^\prime(x_1) + c_2\phi_2f^\prime(x_2) - cf(x_1) - (1 - c)f(x_2) + f(x_3)}{\phi_1g(x_1) + \phi_2g(x_2) +H_0}\\
&- \frac{\phi_1f(x_1) + \phi_2f(x_2)}{(\phi_1g(x_1) + \phi_2g(x_2) + H_0)^2} \bigg(c_1\phi_1g^\prime(x_1) +c_2\phi_2g^\prime(x_2)\\
& - cg(x_1) - (1-c)g(x_2) + g(x_3) + \frac{H_3}{2H_0}\bigg)\bigg] \bigg\},
\end{aligned}
\label{eq:freediff_bin_cc1c2}
\end{equation}
where
\begin{equation}
\begin{aligned}
H_3 =& 2(\phi_1g(x_1) + \phi_2g(x_2))(c_1\phi_1g^\prime(x_1) + c_2\phi_2g^\prime(x_2) -cg(x_1) -(1-c)g(x_2) + g(x_3))\\
&+ D(2c_1\phi_1g(x_1)g^\prime(x_1) + 2c_2\phi_2g(x_2)g^\prime(x_2) - cg^2(x_1) - (1-c)g^2(x_2) + g^2(x_3)).
\end{aligned}
\end{equation}

Since $\phi_1$, $\phi_2$, $x_1$, and $x_2$ are determined by minimizing the two-phase free energy, Eqs.~(\ref{eq:dF_dphi2}) and (\ref{eq:dF_dx2}) should be satisfied when inserting the numerically determined values. This means that we can
substitute Eqs.~(\ref{eq:dF_dphi2}) and (\ref{eq:dF_dx2}) in Eq.~(\ref{eq:freediff_bin_cc1c2}) to eliminate the constants $c_1$, $c_2$, and $c$. This leads to
\begin{equation}
\begin{aligned}
\delta  \Delta F^\star = & \delta \phi \bigg\{\Delta F(x_3) - \Delta F(x_1) + (x_1 - x_3) F^\prime(x_1)\\
&+ 2G\tau\bigg[\frac{f(x_3) -f(x_1) + (x_1 -x_3)f^\prime(x_1)}{\phi_1g(x_1) + \phi_2g(x_2) + H_0}\\
&- \frac{\phi_1f(x_1) + \phi_2f(x_2)}{(\phi_1g(x_1) + \phi_2g(x_2) + H_0)^2} \bigg(g(x_3) - g(x_1) + (x_1 - x_3)g^\prime(x_1)\\
&+ \big[2(\phi_1g(x_1) + \phi_2g(x_2))(g(x_3) -g(x_1) + (x_1 - x_3)g^\prime(x_1)) \\
&+ D(g^2(x_3) -g^2(x_1) + 2(x_1-x_3)g(x_1)g^\prime(x_1))\big]/{2H_0}\bigg) \bigg]\bigg\}.
\end{aligned}
\end{equation}
When $\delta \Delta F^\star < 0$, the three-phase coexistence has a lower free energy than the two-phase coexistence. We thus find the binodals for two-phase to three-phase coexistence in Figs.~\ref{fig:fig2}(d)-(i) of the main text by looking if a value $x_3$ of a third phase exists for which $\delta \Delta F^\star = 0$. Because we have two types of two-phase coexistence we find two types of binodals. The binodals labeled `1' are those for `nearly I-pure + mixed' to three-phase coexistence and the binodals labeled `2' are those for `I-rich + Br-rich' to three-phase coexistence.

To find the spinodals we consider the possibility to decrease the free energy by generating a small volume fraction $\delta \phi_3$ of a third phase from the first phase with a slightly different Br concentration $x_3 = x_1 + \delta x_3$ (the argument is the same for generation of the third phase from the second phase). The free energy for the three-phase coexistence can be written as $\Delta F^\star(x_1 + \delta x_1,  x_2 + \delta x_2, x_1 + \delta x_3, \phi_1 + \delta \phi_1, \phi_2 + \delta \phi_2, \delta \phi_3,T)$ with
\begin{equation}
\delta \phi_1 + \delta \phi_2 + \delta \phi_3 = 0,
\end{equation}
and
\begin{equation}
x_1 \delta \phi_1 + x_2 \delta \phi_2 + x_1 \delta \phi_3 + \phi_1 \delta x_1 + \phi_2 \delta x_2  + \delta \phi_1 \delta x_1 + \delta \phi_2 \delta x_2 + \delta \phi_3 \delta x_3 = 0.
\end{equation}
Taking $x_3 = x_1 + \delta x_3$, the free energy difference between the three-phase and two-phase coexistence is to second order in $\delta x_3$:
\begin{equation}
\begin{aligned}
\delta \Delta F^\star =& \Delta F(x_1) \delta\phi_1 + \Delta F(x_2)\delta\phi_2  +  \Delta F(x_1) \delta\phi_3 + \phi_1 \Delta F^\prime(x_1)\delta x_1 + \phi_2\Delta F^\prime(x_2)\delta x_2\\
&+ \Delta F^\prime(x_1)\delta \phi_1 \delta x_1 + \Delta F^\prime(x_2)\delta \phi_2 \delta x_2 + \Delta F^\prime(x_1)\delta \phi_3 \delta x_3 + \Delta F^{\prime\prime}(x_1)\delta \phi_3\left(\delta x_3\right)^2/2\\
&+2G\tau\bigg[\frac{1}{\phi_1g(x_1) + \phi_2g(x_2) + H_0}\bigg(f(x_1)\delta\phi_1 + f(x_2)\delta\phi_2 + f(x_1)\delta\phi_3\\
& + \phi_1f^\prime(x_1)\delta x_1 + \phi_2f^\prime(x_2)\delta x_2+ f^\prime(x_1)\delta\phi_1\delta x_1 + f^\prime(x_2)\delta\phi_2\delta x_2 + f^\prime(x_1)\delta\phi_3\delta x_3 \\
&+ f^{\prime\prime}(x_1)\delta \phi_3(\delta x_3)^2/2\bigg)-\frac{\phi_1 f(x_1) + \phi_2 f(x_2)}{(\phi_1 g(x_1) + \phi_2 g(x_2) + H_0)^2} \bigg(g(x_1)
\delta\phi_1 + g(x_2)\delta\phi_2 \\
&+ g(x_1)\delta\phi_3 + \phi_1g^\prime(x_1)\delta x_1 + \phi_2g^\prime(x_2)\delta x_2+ g^\prime(x_1)\delta\phi_1\delta x_1 + g^\prime(x_2)\delta\phi_2\delta x_2 \\
& + g^\prime(x_1)\delta\phi_3\delta x_3 + g^{\prime\prime}(x_1)\delta \phi_3(\delta x_3)^2/2 + \frac{H_4}{2H_0}
\bigg)
\bigg],
\end{aligned}
\label{eq:3phasespin0}
\end{equation}
where
\begin{equation}
\begin{aligned}
H_4 =& 2(\phi_1g(x_1) + \phi_2g(x_2))(g(x_1)
\delta\phi_1 + g(x_2)\delta\phi_2 + g(x_1)\delta\phi_3 + \phi_1g^\prime(x_1)\delta x_1 + \phi_2g^\prime(x_2)\delta x_2\\
&+ g^\prime(x_1)\delta\phi_1\delta x_1 + g^\prime(x_2)\delta\phi_2\delta x_2 + g^\prime(x_1)\delta\phi_3\delta x_3 + g^{\prime\prime}(x_1)\delta \phi_3(\delta x_3)^2/2)\\
&+ D[g^2(x_1)\delta\phi_1 + g^2(x_2)\delta\phi_2 + g^2(x_1)\delta\phi_3 + 2(\phi_1g(x_1)g^\prime(x_1)\delta x_1 + \phi_2g(x_2)g^\prime(x_2)\delta x_2\\
&+ g(x_1)g^\prime(x_1)\delta\phi_1\delta x_1 + g(x_2)g^\prime(x_2)\delta\phi_2\delta x_2 + g(x_1)g^\prime(x_1)\delta\phi_3\delta x_3 \\
&+ ([g^\prime(x_1)]^2 + g(x_1)g^{\prime\prime}(x_1))\delta\phi_3(\delta x_3)^2/2)].
\end{aligned}
\end{equation}
Substituting Eqs.~(\ref{eq:dF_dphi2}) and (\ref{eq:dF_dx2}) in Eq.~(\ref{eq:3phasespin0}), the difference in free energy becomes
\begin{equation}
\begin{aligned}
\delta \Delta F^\star = &\frac{1}{2} \delta \phi_3 (\delta x_3)^2\bigg\{\Delta F^{\prime\prime}(x_1) + 2G\tau\bigg[\frac{1}{\phi_1 g(x_1) + \phi_2 g(x_2) + H_0}f^{\prime\prime}(x_1) \\
& - \frac{\phi_1 f(x_1) + \phi_2 f(x_2)}{(\phi_1 g(x_1) + \phi_2 g(x_2) + H_0)^2} \bigg(g^{\prime\prime}(x_1) \\
&+ \frac{(\phi_1g(x_1) + \phi_2g(x_2))g^{\prime\prime}(x_1) + D[(g^\prime(x_1))^2 + g(x_1)g^{\prime\prime}(x_1)]}{H_0}\bigg) \bigg] \bigg\}.
\end{aligned}
\label{eq:3phasespinodal1}
\end{equation}
Starting in the `nearly I-pure + mixed' coexistence region and decreasing $T$ we first cross binodal `1'. After that, the three-phase coexistence free energy becomes lower than the two-phase coexistence free energy, while a barrier between these free energies still exists. Taking $x_1$ as the Br concentration in the mixed phase, the barrier disappears when  $\delta \Delta F^\star =  0$, which is where spinodal `1' is located. Taking instead  $x_1$ as the Br concentration in the nearly I-pure phase, we have $\delta \Delta F^\star >  0$, demonstrating that the I-rich and Br-rich phase are generated from the mixed phase and not from the nearly I-pure phase. Vice versa, starting in the `I-rich + Br-rich' coexistence region and increasing $T$ we first cross binodal `2'. Taking $x_1$ as the Br concentration in the I-rich phase, the barrier disappears when  $\delta \Delta F^\star =  0$, which is where spinodal `2' is located. Taking instead  $x_1$ as the Br concentration in the Br-rich phase, we have $\delta \Delta F^\star >  0$, demonstrating that the nearly I-pure phase is generated from the I-rich phase and not from the Br-rich phase.
The results in Fig.~\ref{fig:fig3} of the main text have been obtained with a standard numerical minimization algorithm.


\begin{thebibliography}{30}%
\makeatletter
\providecommand \@ifxundefined [1]{%
 \@ifx{#1\undefined}
}%
\providecommand \@ifnum [1]{%
 \ifnum #1\expandafter \@firstoftwo
 \else \expandafter \@secondoftwo
 \fi
}%
\providecommand \@ifx [1]{%
 \ifx #1\expandafter \@firstoftwo
 \else \expandafter \@secondoftwo
 \fi
}%
\providecommand \natexlab [1]{#1}%
\providecommand \enquote  [1]{``#1''}%
\providecommand \bibnamefont  [1]{#1}%
\providecommand \bibfnamefont [1]{#1}%
\providecommand \citenamefont [1]{#1}%
\providecommand \href@noop [0]{\@secondoftwo}%
\providecommand \href [0]{\begingroup \@sanitize@url \@href}%
\providecommand \@href[1]{\@@startlink{#1}\@@href}%
\providecommand \@@href[1]{\endgroup#1\@@endlink}%
\providecommand \@sanitize@url [0]{\catcode `\\12\catcode `\$12\catcode
  `\&12\catcode `\#12\catcode `\^12\catcode `\_12\catcode `\%12\relax}%
\providecommand \@@startlink[1]{}%
\providecommand \@@endlink[0]{}%
\providecommand \url  [0]{\begingroup\@sanitize@url \@url }%
\providecommand \@url [1]{\endgroup\@href {#1}{\urlprefix }}%
\providecommand \urlprefix  [0]{URL }%
\providecommand \Eprint [0]{\href }%
\providecommand \doibase [0]{https://doi.org/}%
\providecommand \selectlanguage [0]{\@gobble}%
\providecommand \bibinfo  [0]{\@secondoftwo}%
\providecommand \bibfield  [0]{\@secondoftwo}%
\providecommand \translation [1]{[#1]}%
\providecommand \BibitemOpen [0]{}%
\providecommand \bibitemStop [0]{}%
\providecommand \bibitemNoStop [0]{.\EOS\space}%
\providecommand \EOS [0]{\spacefactor3000\relax}%
\providecommand \BibitemShut  [1]{\csname bibitem#1\endcsname}%
\let\auto@bib@innerbib\@empty
\bibitem [{\citenamefont {Kojima}\ \emph {et~al.}(2009)\citenamefont {Kojima},
  \citenamefont {Teshima}, \citenamefont {Shirai},\ and\ \citenamefont
  {Miyasaka}}]{Kojima2009}%
  \BibitemOpen
  \bibfield  {author} {\bibinfo {author} {\bibfnamefont {A.}~\bibnamefont
  {Kojima}}, \bibinfo {author} {\bibfnamefont {K.}~\bibnamefont {Teshima}},
  \bibinfo {author} {\bibfnamefont {Y.}~\bibnamefont {Shirai}},\ and\ \bibinfo
  {author} {\bibfnamefont {T.}~\bibnamefont {Miyasaka}},\ }\href
  {https://doi.org/10.1021/ja809598r} {\bibfield  {journal} {\bibinfo
  {journal} {J. Am. Chem. Soc.}\ }\textbf {\bibinfo {volume} {131}},\ \bibinfo
  {pages} {6050} (\bibinfo {year} {2009})}\BibitemShut {NoStop}%
\bibitem [{\citenamefont {Green}\ \emph {et~al.}(2014)\citenamefont {Green},
  \citenamefont {Ho-Baillie},\ and\ \citenamefont {Snaith}}]{Green2014}%
  \BibitemOpen
  \bibfield  {author} {\bibinfo {author} {\bibfnamefont {M.~A.}\ \bibnamefont
  {Green}}, \bibinfo {author} {\bibfnamefont {A.}~\bibnamefont {Ho-Baillie}},\
  and\ \bibinfo {author} {\bibfnamefont {H.~J.}\ \bibnamefont {Snaith}},\
  }\href {https://doi.org/10.1038/nphoton.2014.134} {\bibfield  {journal}
  {\bibinfo  {journal} {Nat. Photonics}\ }\textbf {\bibinfo {volume} {8}},\
  \bibinfo {pages} {506} (\bibinfo {year} {2014})}\BibitemShut {NoStop}%
\bibitem [{\citenamefont {Noel}\ \emph {et~al.}(2014)\citenamefont {Noel},
  \citenamefont {Stranks}, \citenamefont {Abate}, \citenamefont {Wehrenfennig},
  \citenamefont {Guarnera}, \citenamefont {Haghighirad}, \citenamefont
  {Sadhanala}, \citenamefont {Eperon}, \citenamefont {Pathak}, \citenamefont
  {Johnston}, \citenamefont {Petrozza}, \citenamefont {Herz},\ and\
  \citenamefont {Snaith}}]{Noel2014}%
  \BibitemOpen
  \bibfield  {author} {\bibinfo {author} {\bibfnamefont {N.~K.}\ \bibnamefont
  {Noel}}, \bibinfo {author} {\bibfnamefont {S.~D.}\ \bibnamefont {Stranks}},
  \bibinfo {author} {\bibfnamefont {A.}~\bibnamefont {Abate}}, \bibinfo
  {author} {\bibfnamefont {C.}~\bibnamefont {Wehrenfennig}}, \bibinfo {author}
  {\bibfnamefont {S.}~\bibnamefont {Guarnera}}, \bibinfo {author}
  {\bibfnamefont {A.~A.}\ \bibnamefont {Haghighirad}}, \bibinfo {author}
  {\bibfnamefont {A.}~\bibnamefont {Sadhanala}}, \bibinfo {author}
  {\bibfnamefont {G.~E.}\ \bibnamefont {Eperon}}, \bibinfo {author}
  {\bibfnamefont {S.~K.}\ \bibnamefont {Pathak}}, \bibinfo {author}
  {\bibfnamefont {M.~B.}\ \bibnamefont {Johnston}}, \bibinfo {author}
  {\bibfnamefont {A.}~\bibnamefont {Petrozza}}, \bibinfo {author}
  {\bibfnamefont {L.~M.}\ \bibnamefont {Herz}},\ and\ \bibinfo {author}
  {\bibfnamefont {H.~J.}\ \bibnamefont {Snaith}},\ }\href
  {https://doi.org/10.1039/c4ee01076k} {\bibfield  {journal} {\bibinfo
  {journal} {Energy Environ. Sci.}\ }\textbf {\bibinfo {volume} {7}},\ \bibinfo
  {pages} {3061} (\bibinfo {year} {2014})}\BibitemShut {NoStop}%
\bibitem [{\citenamefont {Stranks}\ and\ \citenamefont
  {Snaith}(2015)}]{Stranks2015}%
  \BibitemOpen
  \bibfield  {author} {\bibinfo {author} {\bibfnamefont {S.~D.}\ \bibnamefont
  {Stranks}}\ and\ \bibinfo {author} {\bibfnamefont {H.~J.}\ \bibnamefont
  {Snaith}},\ }\href {https://doi.org/10.1038/nnano.2015.90} {\bibfield
  {journal} {\bibinfo  {journal} {Nat. Nanotechnol.}\ }\textbf {\bibinfo
  {volume} {10}},\ \bibinfo {pages} {391} (\bibinfo {year} {2015})}\BibitemShut
  {NoStop}%
\bibitem [{\citenamefont {Jeon}\ \emph {et~al.}(2015)\citenamefont {Jeon},
  \citenamefont {Noh}, \citenamefont {Yang}, \citenamefont {Kim}, \citenamefont
  {Ryu}, \citenamefont {Seo},\ and\ \citenamefont {Seok}}]{Jeon2015}%
  \BibitemOpen
  \bibfield  {author} {\bibinfo {author} {\bibfnamefont {N.~J.}\ \bibnamefont
  {Jeon}}, \bibinfo {author} {\bibfnamefont {J.~H.}\ \bibnamefont {Noh}},
  \bibinfo {author} {\bibfnamefont {W.~S.}\ \bibnamefont {Yang}}, \bibinfo
  {author} {\bibfnamefont {Y.~C.}\ \bibnamefont {Kim}}, \bibinfo {author}
  {\bibfnamefont {S.}~\bibnamefont {Ryu}}, \bibinfo {author} {\bibfnamefont
  {J.}~\bibnamefont {Seo}},\ and\ \bibinfo {author} {\bibfnamefont {S.~I.}\
  \bibnamefont {Seok}},\ }\href {https://doi.org/10.1038/nature14133}
  {\bibfield  {journal} {\bibinfo  {journal} {Nature}\ }\textbf {\bibinfo
  {volume} {517}},\ \bibinfo {pages} {476} (\bibinfo {year}
  {2015})}\BibitemShut {NoStop}%
\bibitem [{\citenamefont {McMeekin}\ \emph {et~al.}(2016)\citenamefont
  {McMeekin}, \citenamefont {Sadoughi}, \citenamefont {Rehman}, \citenamefont
  {Eperon}, \citenamefont {Saliba}, \citenamefont {H{\"{o}}rantner},
  \citenamefont {Haghighirad}, \citenamefont {Sakai}, \citenamefont {Korte},
  \citenamefont {Rech}, \citenamefont {Johnston}, \citenamefont {Herz},\ and\
  \citenamefont {Snaith}}]{McMeekin2016}%
  \BibitemOpen
  \bibfield  {author} {\bibinfo {author} {\bibfnamefont {D.~P.}\ \bibnamefont
  {McMeekin}}, \bibinfo {author} {\bibfnamefont {G.}~\bibnamefont {Sadoughi}},
  \bibinfo {author} {\bibfnamefont {W.}~\bibnamefont {Rehman}}, \bibinfo
  {author} {\bibfnamefont {G.~E.}\ \bibnamefont {Eperon}}, \bibinfo {author}
  {\bibfnamefont {M.}~\bibnamefont {Saliba}}, \bibinfo {author} {\bibfnamefont
  {M.~T.}\ \bibnamefont {H{\"{o}}rantner}}, \bibinfo {author} {\bibfnamefont
  {A.}~\bibnamefont {Haghighirad}}, \bibinfo {author} {\bibfnamefont
  {N.}~\bibnamefont {Sakai}}, \bibinfo {author} {\bibfnamefont
  {L.}~\bibnamefont {Korte}}, \bibinfo {author} {\bibfnamefont
  {B.}~\bibnamefont {Rech}}, \bibinfo {author} {\bibfnamefont {M.~B.}\
  \bibnamefont {Johnston}}, \bibinfo {author} {\bibfnamefont {L.~M.}\
  \bibnamefont {Herz}},\ and\ \bibinfo {author} {\bibfnamefont {H.~J.}\
  \bibnamefont {Snaith}},\ }\href {https://doi.org/10.1126/science.aad5845}
  {\bibfield  {journal} {\bibinfo  {journal} {Science}\ }\textbf {\bibinfo
  {volume} {351}},\ \bibinfo {pages} {151} (\bibinfo {year}
  {2016})}\BibitemShut {NoStop}%
\bibitem [{\citenamefont {Lin}\ \emph {et~al.}(2018)\citenamefont {Lin},
  \citenamefont {Xing}, \citenamefont {Quan}, \citenamefont {de~Arquer},
  \citenamefont {Gong}, \citenamefont {Lu}, \citenamefont {Xie}, \citenamefont
  {Zhao}, \citenamefont {Zhang}, \citenamefont {Yan}, \citenamefont {Li},
  \citenamefont {Liu}, \citenamefont {Lu}, \citenamefont {Kirman},
  \citenamefont {Sargent}, \citenamefont {Xiong},\ and\ \citenamefont
  {Wei}}]{Lin2018}%
  \BibitemOpen
  \bibfield  {author} {\bibinfo {author} {\bibfnamefont {K.}~\bibnamefont
  {Lin}}, \bibinfo {author} {\bibfnamefont {J.}~\bibnamefont {Xing}}, \bibinfo
  {author} {\bibfnamefont {L.~N.}\ \bibnamefont {Quan}}, \bibinfo {author}
  {\bibfnamefont {F.~P.~G.}\ \bibnamefont {de~Arquer}}, \bibinfo {author}
  {\bibfnamefont {X.}~\bibnamefont {Gong}}, \bibinfo {author} {\bibfnamefont
  {J.}~\bibnamefont {Lu}}, \bibinfo {author} {\bibfnamefont {L.}~\bibnamefont
  {Xie}}, \bibinfo {author} {\bibfnamefont {W.}~\bibnamefont {Zhao}}, \bibinfo
  {author} {\bibfnamefont {D.}~\bibnamefont {Zhang}}, \bibinfo {author}
  {\bibfnamefont {C.}~\bibnamefont {Yan}}, \bibinfo {author} {\bibfnamefont
  {W.}~\bibnamefont {Li}}, \bibinfo {author} {\bibfnamefont {X.}~\bibnamefont
  {Liu}}, \bibinfo {author} {\bibfnamefont {Y.}~\bibnamefont {Lu}}, \bibinfo
  {author} {\bibfnamefont {J.}~\bibnamefont {Kirman}}, \bibinfo {author}
  {\bibfnamefont {E.~H.}\ \bibnamefont {Sargent}}, \bibinfo {author}
  {\bibfnamefont {Q.}~\bibnamefont {Xiong}},\ and\ \bibinfo {author}
  {\bibfnamefont {Z.}~\bibnamefont {Wei}},\ }\href
  {https://doi.org/10.1038/s41586-018-0575-3} {\bibfield  {journal} {\bibinfo
  {journal} {Nature}\ }\textbf {\bibinfo {volume} {562}},\ \bibinfo {pages}
  {245} (\bibinfo {year} {2018})}\BibitemShut {NoStop}%
\bibitem [{\citenamefont {Jung}\ \emph {et~al.}(2019)\citenamefont {Jung},
  \citenamefont {Jeon}, \citenamefont {Park}, \citenamefont {Moon},
  \citenamefont {Shin}, \citenamefont {Yang}, \citenamefont {Noh},\ and\
  \citenamefont {Seo}}]{Jung2019}%
  \BibitemOpen
  \bibfield  {author} {\bibinfo {author} {\bibfnamefont {E.~H.}\ \bibnamefont
  {Jung}}, \bibinfo {author} {\bibfnamefont {N.~J.}\ \bibnamefont {Jeon}},
  \bibinfo {author} {\bibfnamefont {E.~Y.}\ \bibnamefont {Park}}, \bibinfo
  {author} {\bibfnamefont {C.~S.}\ \bibnamefont {Moon}}, \bibinfo {author}
  {\bibfnamefont {T.~J.}\ \bibnamefont {Shin}}, \bibinfo {author}
  {\bibfnamefont {T.~Y.}\ \bibnamefont {Yang}}, \bibinfo {author}
  {\bibfnamefont {J.~H.}\ \bibnamefont {Noh}},\ and\ \bibinfo {author}
  {\bibfnamefont {J.}~\bibnamefont {Seo}},\ }\href
  {https://doi.org/10.1038/s41586-019-1036-3} {\bibfield  {journal} {\bibinfo
  {journal} {Nature}\ }\textbf {\bibinfo {volume} {567}},\ \bibinfo {pages}
  {511} (\bibinfo {year} {2019})}\BibitemShut {NoStop}%
\bibitem [{\citenamefont {Al-Ashouri}\ \emph {et~al.}(2020)\citenamefont
  {Al-Ashouri}, \citenamefont {Köhnen}, \citenamefont {Li}, \citenamefont
  {Magomedov}, \citenamefont {Hempel}, \citenamefont {Caprioglio},
  \citenamefont {Márquez}, \citenamefont {Vilches}, \citenamefont
  {Kasparavicius}, \citenamefont {Smith}, \citenamefont {Phung}, \citenamefont
  {Menzel}, \citenamefont {Grischek}, \citenamefont {Kegelmann}, \citenamefont
  {Skroblin}, \citenamefont {Gollwitzer}, \citenamefont {Malinauskas},
  \citenamefont {Jošt}, \citenamefont {Matič}, \citenamefont {Rech},
  \citenamefont {Schlatmann}, \citenamefont {Topič}, \citenamefont {Korte},
  \citenamefont {Abate}, \citenamefont {Stannowski}, \citenamefont {Neher},
  \citenamefont {Stolterfoht}, \citenamefont {Unold}, \citenamefont
  {Getautis},\ and\ \citenamefont {Albrecht}}]{AlAshouri2020}%
  \BibitemOpen
  \bibfield  {author} {\bibinfo {author} {\bibfnamefont {A.}~\bibnamefont
  {Al-Ashouri}}, \bibinfo {author} {\bibfnamefont {E.}~\bibnamefont {Köhnen}},
  \bibinfo {author} {\bibfnamefont {B.}~\bibnamefont {Li}}, \bibinfo {author}
  {\bibfnamefont {A.}~\bibnamefont {Magomedov}}, \bibinfo {author}
  {\bibfnamefont {H.}~\bibnamefont {Hempel}}, \bibinfo {author} {\bibfnamefont
  {P.}~\bibnamefont {Caprioglio}}, \bibinfo {author} {\bibfnamefont {J.~A.}\
  \bibnamefont {Márquez}}, \bibinfo {author} {\bibfnamefont {A.~B.~M.}\
  \bibnamefont {Vilches}}, \bibinfo {author} {\bibfnamefont {E.}~\bibnamefont
  {Kasparavicius}}, \bibinfo {author} {\bibfnamefont {J.~A.}\ \bibnamefont
  {Smith}}, \bibinfo {author} {\bibfnamefont {N.}~\bibnamefont {Phung}},
  \bibinfo {author} {\bibfnamefont {D.}~\bibnamefont {Menzel}}, \bibinfo
  {author} {\bibfnamefont {M.}~\bibnamefont {Grischek}}, \bibinfo {author}
  {\bibfnamefont {L.}~\bibnamefont {Kegelmann}}, \bibinfo {author}
  {\bibfnamefont {D.}~\bibnamefont {Skroblin}}, \bibinfo {author}
  {\bibfnamefont {C.}~\bibnamefont {Gollwitzer}}, \bibinfo {author}
  {\bibfnamefont {T.}~\bibnamefont {Malinauskas}}, \bibinfo {author}
  {\bibfnamefont {M.}~\bibnamefont {Jošt}}, \bibinfo {author} {\bibfnamefont
  {G.}~\bibnamefont {Matič}}, \bibinfo {author} {\bibfnamefont
  {B.}~\bibnamefont {Rech}}, \bibinfo {author} {\bibfnamefont {R.}~\bibnamefont
  {Schlatmann}}, \bibinfo {author} {\bibfnamefont {M.}~\bibnamefont {Topič}},
  \bibinfo {author} {\bibfnamefont {L.}~\bibnamefont {Korte}}, \bibinfo
  {author} {\bibfnamefont {A.}~\bibnamefont {Abate}}, \bibinfo {author}
  {\bibfnamefont {B.}~\bibnamefont {Stannowski}}, \bibinfo {author}
  {\bibfnamefont {D.}~\bibnamefont {Neher}}, \bibinfo {author} {\bibfnamefont
  {M.}~\bibnamefont {Stolterfoht}}, \bibinfo {author} {\bibfnamefont
  {T.}~\bibnamefont {Unold}}, \bibinfo {author} {\bibfnamefont
  {V.}~\bibnamefont {Getautis}},\ and\ \bibinfo {author} {\bibfnamefont
  {S.}~\bibnamefont {Albrecht}},\ }\href
  {https://doi.org/10.1126/science.abd4016} {\bibfield  {journal} {\bibinfo
  {journal} {Science}\ }\textbf {\bibinfo {volume} {370}},\ \bibinfo {pages}
  {1300} (\bibinfo {year} {2020})}\BibitemShut {NoStop}%
\bibitem [{\citenamefont {Xiao}\ \emph {et~al.}(2020)\citenamefont {Xiao},
  \citenamefont {Wen}, \citenamefont {Han}, \citenamefont {Lin}, \citenamefont
  {Gao}, \citenamefont {Gu}, \citenamefont {Zang}, \citenamefont {Nie},
  \citenamefont {Zhu}, \citenamefont {Xu},\ and\ \citenamefont
  {Tan}}]{Xiao2020}%
  \BibitemOpen
  \bibfield  {author} {\bibinfo {author} {\bibfnamefont {K.}~\bibnamefont
  {Xiao}}, \bibinfo {author} {\bibfnamefont {J.}~\bibnamefont {Wen}}, \bibinfo
  {author} {\bibfnamefont {Q.}~\bibnamefont {Han}}, \bibinfo {author}
  {\bibfnamefont {R.}~\bibnamefont {Lin}}, \bibinfo {author} {\bibfnamefont
  {Y.}~\bibnamefont {Gao}}, \bibinfo {author} {\bibfnamefont {S.}~\bibnamefont
  {Gu}}, \bibinfo {author} {\bibfnamefont {Y.}~\bibnamefont {Zang}}, \bibinfo
  {author} {\bibfnamefont {Y.}~\bibnamefont {Nie}}, \bibinfo {author}
  {\bibfnamefont {J.}~\bibnamefont {Zhu}}, \bibinfo {author} {\bibfnamefont
  {J.}~\bibnamefont {Xu}},\ and\ \bibinfo {author} {\bibfnamefont
  {H.}~\bibnamefont {Tan}},\ }\href
  {https://doi.org/10.1021/acsenergylett.0c01184} {\bibfield  {journal}
  {\bibinfo  {journal} {ACS Energy Lett.}\ }\textbf {\bibinfo {volume} {5}},\
  \bibinfo {pages} {2819} (\bibinfo {year} {2020})}\BibitemShut {NoStop}%
\bibitem [{\citenamefont {Wang}\ \emph {et~al.}(2020)\citenamefont {Wang},
  \citenamefont {Zardetto}, \citenamefont {Datta}, \citenamefont {Zhang},
  \citenamefont {Wienk},\ and\ \citenamefont {Janssen}}]{Wang2020}%
  \BibitemOpen
  \bibfield  {author} {\bibinfo {author} {\bibfnamefont {J.}~\bibnamefont
  {Wang}}, \bibinfo {author} {\bibfnamefont {V.}~\bibnamefont {Zardetto}},
  \bibinfo {author} {\bibfnamefont {K.}~\bibnamefont {Datta}}, \bibinfo
  {author} {\bibfnamefont {D.}~\bibnamefont {Zhang}}, \bibinfo {author}
  {\bibfnamefont {M.~M.}\ \bibnamefont {Wienk}},\ and\ \bibinfo {author}
  {\bibfnamefont {R.~A.~J.}\ \bibnamefont {Janssen}},\ }\href
  {https://doi.org/10.1038/s41467-020-19062-8} {\bibfield  {journal} {\bibinfo
  {journal} {Nat. Commun.}\ }\textbf {\bibinfo {volume} {11}},\ \bibinfo
  {pages} {5254} (\bibinfo {year} {2020})}\BibitemShut {NoStop}%
\bibitem [{\citenamefont {Fu}\ \emph {et~al.}(2022)\citenamefont {Fu},
  \citenamefont {Li}, \citenamefont {Yang}, \citenamefont {Liang},
  \citenamefont {Faes}, \citenamefont {Jeangros}, \citenamefont {Ballif},\ and\
  \citenamefont {Hou}}]{Fu2022}%
  \BibitemOpen
  \bibfield  {author} {\bibinfo {author} {\bibfnamefont {F.}~\bibnamefont
  {Fu}}, \bibinfo {author} {\bibfnamefont {J.}~\bibnamefont {Li}}, \bibinfo
  {author} {\bibfnamefont {T.~C.-J.}\ \bibnamefont {Yang}}, \bibinfo {author}
  {\bibfnamefont {H.}~\bibnamefont {Liang}}, \bibinfo {author} {\bibfnamefont
  {A.}~\bibnamefont {Faes}}, \bibinfo {author} {\bibfnamefont {Q.}~\bibnamefont
  {Jeangros}}, \bibinfo {author} {\bibfnamefont {C.}~\bibnamefont {Ballif}},\
  and\ \bibinfo {author} {\bibfnamefont {Y.}~\bibnamefont {Hou}},\ }\href@noop
  {} {\bibfield  {journal} {\bibinfo  {journal} {Adv. Mater.}\ }\textbf
  {\bibinfo {volume} {34}},\ \bibinfo {pages} {2106540} (\bibinfo {year}
  {2022})}\BibitemShut {NoStop}%
\bibitem [{\citenamefont {Lehmann}\ \emph {et~al.}(2019)\citenamefont
  {Lehmann}, \citenamefont {Franz}, \citenamefont {Többens}, \citenamefont
  {Levcenco}, \citenamefont {Unold}, \citenamefont {Taubert},\ and\
  \citenamefont {Schorr}}]{Lehmann2019}%
  \BibitemOpen
  \bibfield  {author} {\bibinfo {author} {\bibfnamefont {F.}~\bibnamefont
  {Lehmann}}, \bibinfo {author} {\bibfnamefont {A.}~\bibnamefont {Franz}},
  \bibinfo {author} {\bibfnamefont {D.}~\bibnamefont {Többens}}, \bibinfo
  {author} {\bibfnamefont {S.}~\bibnamefont {Levcenco}}, \bibinfo {author}
  {\bibfnamefont {T.}~\bibnamefont {Unold}}, \bibinfo {author} {\bibfnamefont
  {A.}~\bibnamefont {Taubert}},\ and\ \bibinfo {author} {\bibfnamefont
  {S.}~\bibnamefont {Schorr}},\ }\href@noop {} {\bibfield  {journal} {\bibinfo
  {journal} {RSC Adv.}\ }\textbf {\bibinfo {volume} {9}},\ \bibinfo {pages}
  {11151} (\bibinfo {year} {2019})}\BibitemShut {NoStop}%
\bibitem [{\citenamefont {Brivio}\ \emph {et~al.}(2016)\citenamefont {Brivio},
  \citenamefont {Caetano},\ and\ \citenamefont {Walsh}}]{Brivio2016}%
  \BibitemOpen
  \bibfield  {author} {\bibinfo {author} {\bibfnamefont {F.}~\bibnamefont
  {Brivio}}, \bibinfo {author} {\bibfnamefont {C.}~\bibnamefont {Caetano}},\
  and\ \bibinfo {author} {\bibfnamefont {A.}~\bibnamefont {Walsh}},\ }\href
  {https://doi.org/10.1021/acs.jpclett.6b00226} {\bibfield  {journal} {\bibinfo
   {journal} {J. Phys. Chem. Lett.}\ }\textbf {\bibinfo {volume} {7}},\
  \bibinfo {pages} {1083} (\bibinfo {year} {2016})}\BibitemShut {NoStop}%
\bibitem [{\citenamefont {Hoke}\ \emph {et~al.}(2015)\citenamefont {Hoke},
  \citenamefont {Slotcavage}, \citenamefont {Dohner}, \citenamefont {Bowring},
  \citenamefont {Karunadasa},\ and\ \citenamefont {McGehee}}]{Hoke2015}%
  \BibitemOpen
  \bibfield  {author} {\bibinfo {author} {\bibfnamefont {E.~T.}\ \bibnamefont
  {Hoke}}, \bibinfo {author} {\bibfnamefont {D.~J.}\ \bibnamefont
  {Slotcavage}}, \bibinfo {author} {\bibfnamefont {E.~R.}\ \bibnamefont
  {Dohner}}, \bibinfo {author} {\bibfnamefont {A.~R.}\ \bibnamefont {Bowring}},
  \bibinfo {author} {\bibfnamefont {H.~I.}\ \bibnamefont {Karunadasa}},\ and\
  \bibinfo {author} {\bibfnamefont {M.~D.}\ \bibnamefont {McGehee}},\ }\href
  {https://doi.org/10.1039/c4sc03141e} {\bibfield  {journal} {\bibinfo
  {journal} {Chem. Sci.}\ }\textbf {\bibinfo {volume} {6}},\ \bibinfo {pages}
  {613} (\bibinfo {year} {2015})}\BibitemShut {NoStop}%
\bibitem [{\citenamefont {Beal}\ \emph {et~al.}(2016)\citenamefont {Beal},
  \citenamefont {Slotcavage}, \citenamefont {Leijtens}, \citenamefont
  {Bowring}, \citenamefont {Belisle}, \citenamefont {Nguyen}, \citenamefont
  {Burkhard}, \citenamefont {Hoke},\ and\ \citenamefont {McGehee}}]{Beal2016}%
  \BibitemOpen
  \bibfield  {author} {\bibinfo {author} {\bibfnamefont {R.~E.}\ \bibnamefont
  {Beal}}, \bibinfo {author} {\bibfnamefont {D.~J.}\ \bibnamefont
  {Slotcavage}}, \bibinfo {author} {\bibfnamefont {T.}~\bibnamefont
  {Leijtens}}, \bibinfo {author} {\bibfnamefont {A.~R.}\ \bibnamefont
  {Bowring}}, \bibinfo {author} {\bibfnamefont {R.~A.}\ \bibnamefont
  {Belisle}}, \bibinfo {author} {\bibfnamefont {W.~H.}\ \bibnamefont {Nguyen}},
  \bibinfo {author} {\bibfnamefont {G.~F.}\ \bibnamefont {Burkhard}}, \bibinfo
  {author} {\bibfnamefont {E.~T.}\ \bibnamefont {Hoke}},\ and\ \bibinfo
  {author} {\bibfnamefont {M.~D.}\ \bibnamefont {McGehee}},\ }\href
  {https://doi.org/10.1021/acs.jpclett.6b00002} {\bibfield  {journal} {\bibinfo
   {journal} {J. Phys. Chem. Lett.}\ }\textbf {\bibinfo {volume} {7}},\
  \bibinfo {pages} {746} (\bibinfo {year} {2016})}\BibitemShut {NoStop}%
\bibitem [{\citenamefont {Draguta}\ \emph {et~al.}(2017)\citenamefont
  {Draguta}, \citenamefont {Sharia}, \citenamefont {Yoon}, \citenamefont
  {Brennan}, \citenamefont {Morozov}, \citenamefont {Manser}, \citenamefont
  {Kamat}, \citenamefont {Schneider},\ and\ \citenamefont
  {Kuno}}]{Draguta2017}%
  \BibitemOpen
  \bibfield  {author} {\bibinfo {author} {\bibfnamefont {S.}~\bibnamefont
  {Draguta}}, \bibinfo {author} {\bibfnamefont {O.}~\bibnamefont {Sharia}},
  \bibinfo {author} {\bibfnamefont {S.~J.}\ \bibnamefont {Yoon}}, \bibinfo
  {author} {\bibfnamefont {M.~C.}\ \bibnamefont {Brennan}}, \bibinfo {author}
  {\bibfnamefont {Y.~V.}\ \bibnamefont {Morozov}}, \bibinfo {author}
  {\bibfnamefont {J.~M.}\ \bibnamefont {Manser}}, \bibinfo {author}
  {\bibfnamefont {P.~V.}\ \bibnamefont {Kamat}}, \bibinfo {author}
  {\bibfnamefont {W.~F.}\ \bibnamefont {Schneider}},\ and\ \bibinfo {author}
  {\bibfnamefont {M.}~\bibnamefont {Kuno}},\ }\href@noop {} {\bibfield
  {journal} {\bibinfo  {journal} {Nat. Commun.}\ }\textbf {\bibinfo {volume}
  {8}} (\bibinfo {year} {2017})}\BibitemShut {NoStop}%
\bibitem [{\citenamefont {Chen}\ \emph {et~al.}(2021)\citenamefont {Chen},
  \citenamefont {Brocks}, \citenamefont {Tao},\ and\ \citenamefont
  {Bobbert}}]{Chen2021}%
  \BibitemOpen
  \bibfield  {author} {\bibinfo {author} {\bibfnamefont {Z.}~\bibnamefont
  {Chen}}, \bibinfo {author} {\bibfnamefont {G.}~\bibnamefont {Brocks}},
  \bibinfo {author} {\bibfnamefont {S.}~\bibnamefont {Tao}},\ and\ \bibinfo
  {author} {\bibfnamefont {P.~A.}\ \bibnamefont {Bobbert}},\ }\href
  {https://doi.org/10.1038/s41467-021-23008-z} {\bibfield  {journal} {\bibinfo
  {journal} {Nat. Commun.}\ }\textbf {\bibinfo {volume} {12}},\ \bibinfo
  {pages} {1} (\bibinfo {year} {2021})}\BibitemShut {NoStop}%
\bibitem [{\citenamefont {Barker}\ \emph {et~al.}(2017)\citenamefont {Barker},
  \citenamefont {Sadhanala}, \citenamefont {Deschler}, \citenamefont {Gandini},
  \citenamefont {Senanayak}, \citenamefont {Pearce}, \citenamefont {Mosconi},
  \citenamefont {Pearson}, \citenamefont {Wu}, \citenamefont {{Srimath
  Kandada}}, \citenamefont {Leijtens}, \citenamefont {{De Angelis}},
  \citenamefont {Dutton}, \citenamefont {Petrozza},\ and\ \citenamefont
  {Friend}}]{barker2017}%
  \BibitemOpen
  \bibfield  {author} {\bibinfo {author} {\bibfnamefont {A.~J.}\ \bibnamefont
  {Barker}}, \bibinfo {author} {\bibfnamefont {A.}~\bibnamefont {Sadhanala}},
  \bibinfo {author} {\bibfnamefont {F.}~\bibnamefont {Deschler}}, \bibinfo
  {author} {\bibfnamefont {M.}~\bibnamefont {Gandini}}, \bibinfo {author}
  {\bibfnamefont {S.~P.}\ \bibnamefont {Senanayak}}, \bibinfo {author}
  {\bibfnamefont {P.~M.}\ \bibnamefont {Pearce}}, \bibinfo {author}
  {\bibfnamefont {E.}~\bibnamefont {Mosconi}}, \bibinfo {author} {\bibfnamefont
  {A.~J.}\ \bibnamefont {Pearson}}, \bibinfo {author} {\bibfnamefont
  {Y.}~\bibnamefont {Wu}}, \bibinfo {author} {\bibfnamefont {A.~R.}\
  \bibnamefont {{Srimath Kandada}}}, \bibinfo {author} {\bibfnamefont
  {T.}~\bibnamefont {Leijtens}}, \bibinfo {author} {\bibfnamefont
  {F.}~\bibnamefont {{De Angelis}}}, \bibinfo {author} {\bibfnamefont {S.~E.}\
  \bibnamefont {Dutton}}, \bibinfo {author} {\bibfnamefont {A.}~\bibnamefont
  {Petrozza}},\ and\ \bibinfo {author} {\bibfnamefont {R.~H.}\ \bibnamefont
  {Friend}},\ }\href {https://doi.org/10.1021/acsenergylett.7b00282} {\bibfield
   {journal} {\bibinfo  {journal} {ACS Energy Lett.}\ }\textbf {\bibinfo
  {volume} {2}},\ \bibinfo {pages} {1416} (\bibinfo {year} {2017})}\BibitemShut
  {NoStop}%
\bibitem [{\citenamefont {Hutter}\ \emph {et~al.}(2020)\citenamefont {Hutter},
  \citenamefont {Muscarella}, \citenamefont {Wittmann}, \citenamefont
  {Versluis}, \citenamefont {McGovern}, \citenamefont {Bakker}, \citenamefont
  {Woo}, \citenamefont {Jung}, \citenamefont {Walsh},\ and\ \citenamefont
  {Ehrler}}]{Hutter2020}%
  \BibitemOpen
  \bibfield  {author} {\bibinfo {author} {\bibfnamefont {E.~M.}\ \bibnamefont
  {Hutter}}, \bibinfo {author} {\bibfnamefont {L.~A.}\ \bibnamefont
  {Muscarella}}, \bibinfo {author} {\bibfnamefont {F.}~\bibnamefont
  {Wittmann}}, \bibinfo {author} {\bibfnamefont {J.}~\bibnamefont {Versluis}},
  \bibinfo {author} {\bibfnamefont {L.}~\bibnamefont {McGovern}}, \bibinfo
  {author} {\bibfnamefont {H.~J.}\ \bibnamefont {Bakker}}, \bibinfo {author}
  {\bibfnamefont {Y.-W.}\ \bibnamefont {Woo}}, \bibinfo {author} {\bibfnamefont
  {Y.-K.}\ \bibnamefont {Jung}}, \bibinfo {author} {\bibfnamefont
  {A.}~\bibnamefont {Walsh}},\ and\ \bibinfo {author} {\bibfnamefont
  {B.}~\bibnamefont {Ehrler}},\ }\href
  {https://doi.org/10.1016/j.xcrp.2020.100120} {\bibfield  {journal} {\bibinfo
  {journal} {Cell Reports Phys. Sci.}\ }\textbf {\bibinfo {volume} {1}},\
  \bibinfo {pages} {100120} (\bibinfo {year} {2020})}\BibitemShut {NoStop}%
\bibitem [{\citenamefont {Braly}\ \emph {et~al.}(2017)\citenamefont {Braly},
  \citenamefont {Stoddard}, \citenamefont {Rajagopal}, \citenamefont {Uhl},
  \citenamefont {Katahara}, \citenamefont {Jen},\ and\ \citenamefont
  {Hillhouse}}]{Braly2017}%
  \BibitemOpen
  \bibfield  {author} {\bibinfo {author} {\bibfnamefont {I.~L.}\ \bibnamefont
  {Braly}}, \bibinfo {author} {\bibfnamefont {R.~J.}\ \bibnamefont {Stoddard}},
  \bibinfo {author} {\bibfnamefont {A.}~\bibnamefont {Rajagopal}}, \bibinfo
  {author} {\bibfnamefont {A.~R.}\ \bibnamefont {Uhl}}, \bibinfo {author}
  {\bibfnamefont {J.~K.}\ \bibnamefont {Katahara}}, \bibinfo {author}
  {\bibfnamefont {A.~K.}\ \bibnamefont {Jen}},\ and\ \bibinfo {author}
  {\bibfnamefont {H.~W.}\ \bibnamefont {Hillhouse}},\ }\href
  {https://doi.org/10.1021/acsenergylett.7b00525} {\bibfield  {journal}
  {\bibinfo  {journal} {ACS Energy Lett.}\ }\textbf {\bibinfo {volume} {2}},\
  \bibinfo {pages} {1841} (\bibinfo {year} {2017})}\BibitemShut {NoStop}%
\bibitem [{\citenamefont {Rehman}\ \emph {et~al.}(2017)\citenamefont {Rehman},
  \citenamefont {McMeekin}, \citenamefont {Patel}, \citenamefont {Milot},
  \citenamefont {Johnston}, \citenamefont {Snaith},\ and\ \citenamefont
  {Herz}}]{Rehman2017}%
  \BibitemOpen
  \bibfield  {author} {\bibinfo {author} {\bibfnamefont {W.}~\bibnamefont
  {Rehman}}, \bibinfo {author} {\bibfnamefont {D.~P.}\ \bibnamefont
  {McMeekin}}, \bibinfo {author} {\bibfnamefont {J.~B.}\ \bibnamefont {Patel}},
  \bibinfo {author} {\bibfnamefont {R.~L.}\ \bibnamefont {Milot}}, \bibinfo
  {author} {\bibfnamefont {M.~B.}\ \bibnamefont {Johnston}}, \bibinfo {author}
  {\bibfnamefont {H.~J.}\ \bibnamefont {Snaith}},\ and\ \bibinfo {author}
  {\bibfnamefont {L.~M.}\ \bibnamefont {Herz}},\ }\href
  {https://doi.org/10.1039/c6ee03014a} {\bibfield  {journal} {\bibinfo
  {journal} {Energy Environ. Sci.}\ }\textbf {\bibinfo {volume} {10}},\
  \bibinfo {pages} {361} (\bibinfo {year} {2017})}\BibitemShut {NoStop}%
\bibitem [{\citenamefont {Dang}\ \emph {et~al.}(2019)\citenamefont {Dang},
  \citenamefont {Wang}, \citenamefont {Ghasemi}, \citenamefont {Tang},
  \citenamefont {{De Bastiani}}, \citenamefont {Aydin}, \citenamefont {Dauzon},
  \citenamefont {Barrit}, \citenamefont {Peng}, \citenamefont {Smilgies},
  \citenamefont {{De Wolf}},\ and\ \citenamefont {Amassian}}]{Dang2019}%
  \BibitemOpen
  \bibfield  {author} {\bibinfo {author} {\bibfnamefont {H.~X.}\ \bibnamefont
  {Dang}}, \bibinfo {author} {\bibfnamefont {K.}~\bibnamefont {Wang}}, \bibinfo
  {author} {\bibfnamefont {M.}~\bibnamefont {Ghasemi}}, \bibinfo {author}
  {\bibfnamefont {M.~C.}\ \bibnamefont {Tang}}, \bibinfo {author}
  {\bibfnamefont {M.}~\bibnamefont {{De Bastiani}}}, \bibinfo {author}
  {\bibfnamefont {E.}~\bibnamefont {Aydin}}, \bibinfo {author} {\bibfnamefont
  {E.}~\bibnamefont {Dauzon}}, \bibinfo {author} {\bibfnamefont
  {D.}~\bibnamefont {Barrit}}, \bibinfo {author} {\bibfnamefont
  {J.}~\bibnamefont {Peng}}, \bibinfo {author} {\bibfnamefont {D.~M.}\
  \bibnamefont {Smilgies}}, \bibinfo {author} {\bibfnamefont {S.}~\bibnamefont
  {{De Wolf}}},\ and\ \bibinfo {author} {\bibfnamefont {A.}~\bibnamefont
  {Amassian}},\ }\href {https://doi.org/10.1016/j.joule.2019.05.016} {\bibfield
   {journal} {\bibinfo  {journal} {Joule}\ }\textbf {\bibinfo {volume} {3}},\
  \bibinfo {pages} {1746} (\bibinfo {year} {2019})}\BibitemShut {NoStop}%
\bibitem [{\citenamefont {Bush}\ \emph {et~al.}(2018)\citenamefont {Bush},
  \citenamefont {Frohna}, \citenamefont {Prasanna}, \citenamefont {Beal},
  \citenamefont {Leijtens}, \citenamefont {Swifter},\ and\ \citenamefont
  {McGehee}}]{bush2018}%
  \BibitemOpen
  \bibfield  {author} {\bibinfo {author} {\bibfnamefont {K.~A.}\ \bibnamefont
  {Bush}}, \bibinfo {author} {\bibfnamefont {K.}~\bibnamefont {Frohna}},
  \bibinfo {author} {\bibfnamefont {R.}~\bibnamefont {Prasanna}}, \bibinfo
  {author} {\bibfnamefont {R.~E.}\ \bibnamefont {Beal}}, \bibinfo {author}
  {\bibfnamefont {T.}~\bibnamefont {Leijtens}}, \bibinfo {author}
  {\bibfnamefont {S.~A.}\ \bibnamefont {Swifter}},\ and\ \bibinfo {author}
  {\bibfnamefont {M.~D.}\ \bibnamefont {McGehee}},\ }\href
  {https://doi.org/10.1021/acsenergylett.7b01255} {\bibfield  {journal}
  {\bibinfo  {journal} {ACS Energy Lett.}\ }\textbf {\bibinfo {volume} {3}},\
  \bibinfo {pages} {428} (\bibinfo {year} {2018})}\BibitemShut {NoStop}%
\bibitem [{\citenamefont {Noh}\ \emph {et~al.}(2013)\citenamefont {Noh},
  \citenamefont {Im}, \citenamefont {Heo}, \citenamefont {Mandal},\ and\
  \citenamefont {Seok}}]{Noh2013}%
  \BibitemOpen
  \bibfield  {author} {\bibinfo {author} {\bibfnamefont {J.~H.}\ \bibnamefont
  {Noh}}, \bibinfo {author} {\bibfnamefont {S.~H.}\ \bibnamefont {Im}},
  \bibinfo {author} {\bibfnamefont {J.~H.}\ \bibnamefont {Heo}}, \bibinfo
  {author} {\bibfnamefont {T.~N.}\ \bibnamefont {Mandal}},\ and\ \bibinfo
  {author} {\bibfnamefont {S.~I.}\ \bibnamefont {Seok}},\ }\href
  {https://doi.org/10.1021/nl400349b} {\bibfield  {journal} {\bibinfo
  {journal} {Nano Lett.}\ }\textbf {\bibinfo {volume} {13}},\ \bibinfo {pages}
  {1764} (\bibinfo {year} {2013})}\BibitemShut {NoStop}%
\bibitem [{\citenamefont {Johnston}\ and\ \citenamefont
  {Herz}(2016)}]{Johnston2016}%
  \BibitemOpen
  \bibfield  {author} {\bibinfo {author} {\bibfnamefont {M.~B.}\ \bibnamefont
  {Johnston}}\ and\ \bibinfo {author} {\bibfnamefont {L.~M.}\ \bibnamefont
  {Herz}},\ }\href {https://doi.org/10.1021/acs.accounts.5b00411} {\bibfield
  {journal} {\bibinfo  {journal} {Acc. Chem. Res.}\ }\textbf {\bibinfo {volume}
  {49}},\ \bibinfo {pages} {146} (\bibinfo {year} {2016})}\BibitemShut
  {NoStop}%
\bibitem [{\citenamefont {Ruth}\ \emph {et~al.}(2018)\citenamefont {Ruth},
  \citenamefont {Brennan}, \citenamefont {Draguta}, \citenamefont {Morozov},
  \citenamefont {Zhukovskyi}, \citenamefont {Janko}, \citenamefont {Zapol},\
  and\ \citenamefont {Kuno}}]{Ruth2018}%
  \BibitemOpen
  \bibfield  {author} {\bibinfo {author} {\bibfnamefont {A.}~\bibnamefont
  {Ruth}}, \bibinfo {author} {\bibfnamefont {M.~C.}\ \bibnamefont {Brennan}},
  \bibinfo {author} {\bibfnamefont {S.}~\bibnamefont {Draguta}}, \bibinfo
  {author} {\bibfnamefont {Y.~V.}\ \bibnamefont {Morozov}}, \bibinfo {author}
  {\bibfnamefont {M.}~\bibnamefont {Zhukovskyi}}, \bibinfo {author}
  {\bibfnamefont {B.}~\bibnamefont {Janko}}, \bibinfo {author} {\bibfnamefont
  {P.}~\bibnamefont {Zapol}},\ and\ \bibinfo {author} {\bibfnamefont
  {M.}~\bibnamefont {Kuno}},\ }\href
  {https://doi.org/10.1021/acsenergylett.8b01369} {\bibfield  {journal}
  {\bibinfo  {journal} {ACS Energy Lett.}\ }\textbf {\bibinfo {volume} {3}},\
  \bibinfo {pages} {2321} (\bibinfo {year} {2018})}\BibitemShut {NoStop}%
\bibitem [{\citenamefont {Brennan}\ \emph {et~al.}(2018)\citenamefont
  {Brennan}, \citenamefont {Draguta}, \citenamefont {Kamat},\ and\
  \citenamefont {Kuno}}]{brennan2018}%
  \BibitemOpen
  \bibfield  {author} {\bibinfo {author} {\bibfnamefont {M.~C.}\ \bibnamefont
  {Brennan}}, \bibinfo {author} {\bibfnamefont {S.}~\bibnamefont {Draguta}},
  \bibinfo {author} {\bibfnamefont {P.~V.}\ \bibnamefont {Kamat}},\ and\
  \bibinfo {author} {\bibfnamefont {M.}~\bibnamefont {Kuno}},\ }\href
  {https://doi.org/10.1021/acsenergylett.7b01151} {\bibfield  {journal}
  {\bibinfo  {journal} {ACS Energy Lett.}\ }\textbf {\bibinfo {volume} {3}},\
  \bibinfo {pages} {204} (\bibinfo {year} {2018})}\BibitemShut {NoStop}%
\bibitem [{\citenamefont {Elmelund}\ \emph {et~al.}(2020)\citenamefont
  {Elmelund}, \citenamefont {Seger}, \citenamefont {Kuno},\ and\ \citenamefont
  {Kamat}}]{Elmelund2020}%
  \BibitemOpen
  \bibfield  {author} {\bibinfo {author} {\bibfnamefont {T.}~\bibnamefont
  {Elmelund}}, \bibinfo {author} {\bibfnamefont {B.}~\bibnamefont {Seger}},
  \bibinfo {author} {\bibfnamefont {M.}~\bibnamefont {Kuno}},\ and\ \bibinfo
  {author} {\bibfnamefont {P.~V.}\ \bibnamefont {Kamat}},\ }\href
  {https://doi.org/10.1021/acsenergylett.9b02265} {\bibfield  {journal}
  {\bibinfo  {journal} {ACS Energy Lett.}\ }\textbf {\bibinfo {volume} {5}},\
  \bibinfo {pages} {56} (\bibinfo {year} {2020})}\BibitemShut {NoStop}%
\bibitem [{\citenamefont {Sher}\ \emph {et~al.}(1987)\citenamefont {Sher},
  \citenamefont {van Schilfgaarde}, \citenamefont {Chen},\ and\ \citenamefont
  {Chen}}]{Sher1987}%
  \BibitemOpen
  \bibfield  {author} {\bibinfo {author} {\bibfnamefont {A.}~\bibnamefont
  {Sher}}, \bibinfo {author} {\bibfnamefont {M.}~\bibnamefont {van
  Schilfgaarde}}, \bibinfo {author} {\bibfnamefont {A.-B.}\ \bibnamefont
  {Chen}},\ and\ \bibinfo {author} {\bibfnamefont {W.}~\bibnamefont {Chen}},\
  }\href {https://doi.org/10.1103/PhysRevB.36.4279} {\bibfield  {journal}
  {\bibinfo  {journal} {Phys. Rev. B}\ }\textbf {\bibinfo {volume} {36}},\
  \bibinfo {pages} {4279} (\bibinfo {year} {1987})}\BibitemShut {NoStop}%
\end{thebibliography}
%

\newpage

\begin{figure}[h]
\centering
\includegraphics[width=0.58\columnwidth]{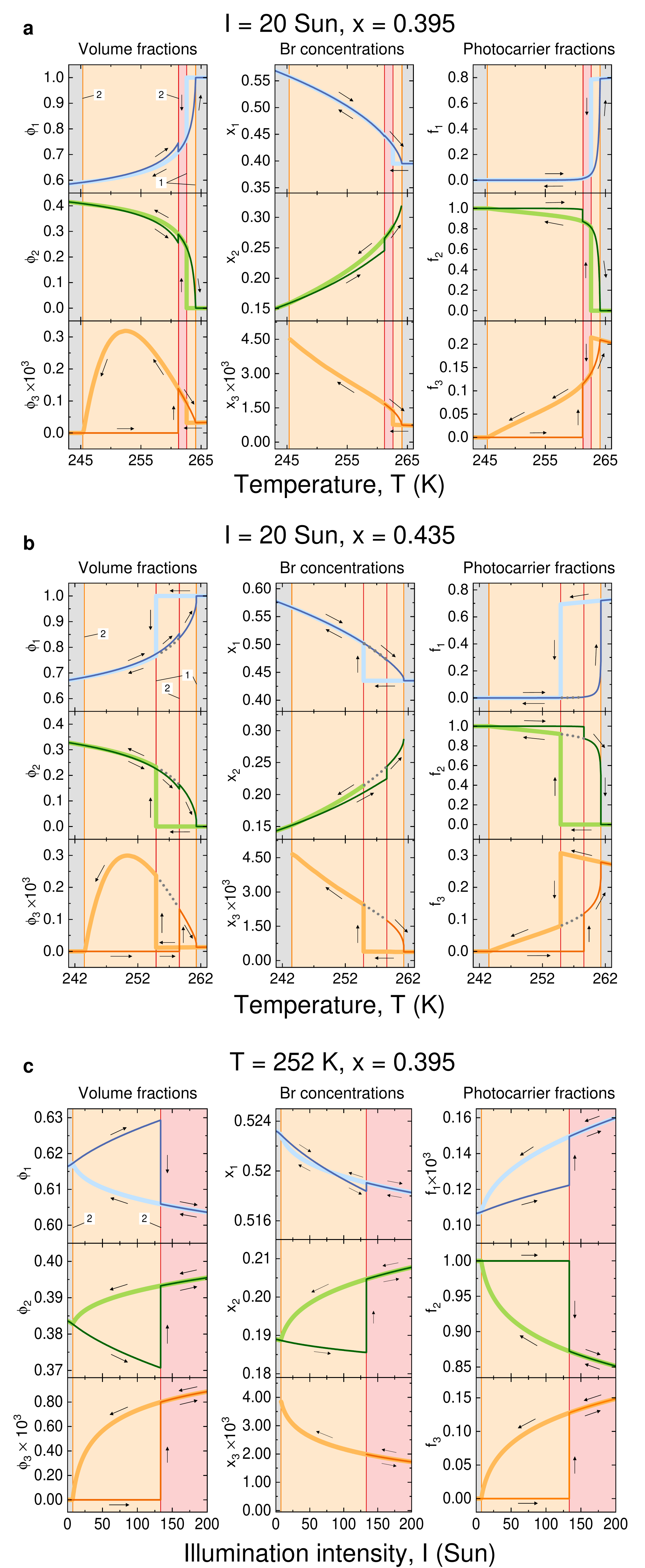}
\end{figure}
\begin{figure}[h]
\centering
\caption{\label{fig:fig3} Parameter sweeps through three-phase coexistence region. Volume fractions $\phi_i$, bromine concentrations $x_i$, and photocarrier fractions $f_i$ as function of temperature $T$ at illumination intensity $I=20$ Sun for (a) $x=0.395$ (left double-headed arrow in Fig.~\ref{fig:fig2}(h)) and (b) $x=0.435$ (right double-headed arrow). From top to bottom: Br-rich phase, I-rich phase, and nearly I-pure phase. Thin dark-colored lines: increasing $T$. Thick light-colored lines: decreasing $T$. Free energy minima are followed until they disappear. The dotted lines in (b) indicate that the three-phase coexistence is separated from the two types of two-phase coexistence by a free energy barrier. (c) Same as a and (b), but as function of $I$ for $T=252$ K and $x=0.395$ (black squares in Figs.~\ref{fig:fig2}(g)-(i)). }
\end{figure}

\clearpage
\begin{figure}[h]
\centering
\includegraphics[width=0.6\columnwidth]{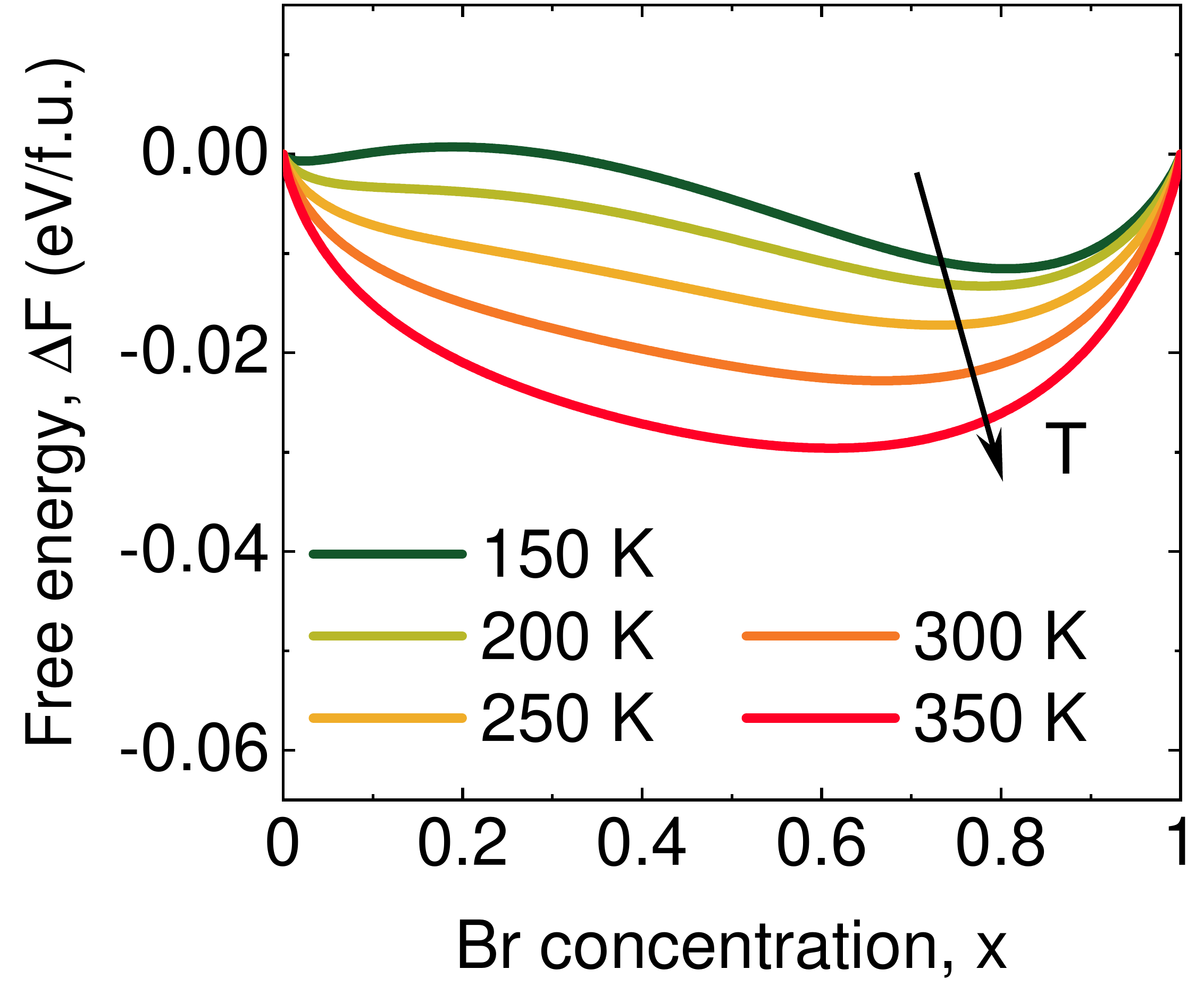}\caption{\label{fig:fig4} Compositional mixing free energy per formula unit (f.u.) as a function of Br concentration $x$ at different temperatures $T$. Reproduced from Ref.~\cite{Chen2021}.}
\label{fig:compositionalfree}
\end{figure}

\clearpage
\begin{figure}[h]
\centering
\includegraphics[width=0.8\columnwidth]{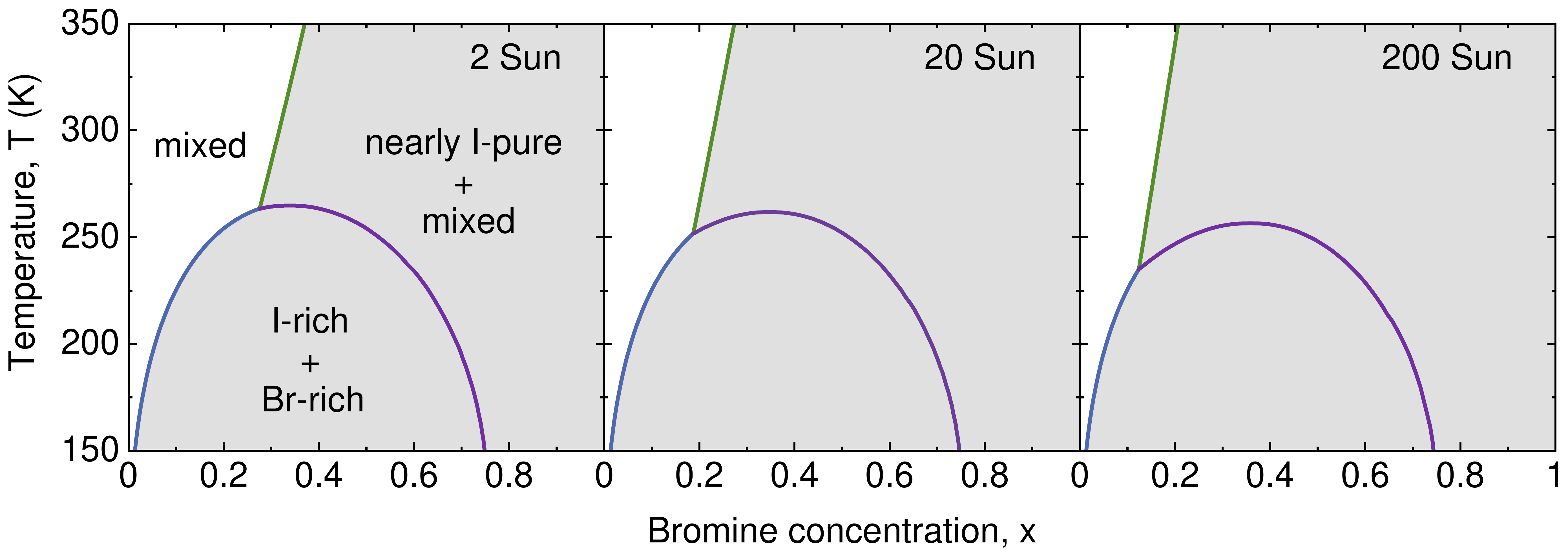}
\caption{Phase diagrams for two-phase coexistence in MAPb(I$_{1-x}$Br$_{x}$)$_3$ at different illumination intensities. The purple lines indicate where the free energy minima of the two different types of two-phase coexistence switch order. }
\label{fig:fig5}
\end{figure}

\end{document}